\documentclass[runningheads]{llncs}

\usepackage{amsmath, amssymb}
\usepackage{microtype}
\usepackage[T1]{fontenc}
\usepackage{lmodern}
\usepackage{textcomp}
\usepackage{latexsym}
\usepackage{bm}
\usepackage{verbatim}
\usepackage{wrapfig}
\usepackage{ascmac}
\usepackage{makeidx}
\usepackage{latexsym}
\usepackage[pdftex]{graphicx}
\usepackage{color}

\begin{document}

\title{Optimal Partition of a Tree with Social Distance\thanks{This work is supported by JSPS KAKENHI Grant Numbers 17K19960, 17H01698, 18H06469.}}




\author{Masahiro Okubo\inst{1}
\and
Tesshu Hanaka\inst{2}
\and
Hirotaka Ono\inst{3}
}

\authorrunning{M. Okubo et al.}

\institute{Graduate School of Informatics, Nagoya University, Furo-cho, Chikusa-ku, Nagoya, Japan \email{okubo.masahiro@h.mbox.nagoya-u.ac.jp} \and
Department of Information and System Engineering, Chuo University, 1-13-27 Kasuga, Bunkyo-ku, Tokyo, Japan \email{hanaka.91t@g.chuo-u.ac.jp} \and Graduate School of Informatics, Nagoya University, Furo-cho, Chikusa-ku, Nagoya, Japan \email{ono@i.nagoya-u.ac.jp}}
\maketitle              












\begin{abstract}
We study the problem to find a partition of \textcolor{black}{a} graph $G$ with maximum social welfare  based on social distance between vertices in $G$, called {\sf MaxSWP}. 
This problem is known to be NP-hard in general. In this paper, we first give a complete characterization of optimal partitions of trees with small diameters. Then, by utilizing these results, we show that {\sf MaxSWP} can be solved in linear time for trees. 
Moreover, we show that {\sf MaxSWP} is NP-hard even for 4-regular graphs.

\keywords{graph algorithm \and tree \and graph partition \and social distance} 
\end{abstract}
\section{Introduction}
With the development of \textcolor{black}{Social Networking Services (SNS)} such as Twitter,
Facebook, Instagram and so on, it has become much easier than before to obtain
graphs that represent human relationship, and there are many attempts
to utilize such graphs for extracting useful information.
Among them, grouping people according to the graph structures is
focused and investigated from many 
standpoints. For example, if a community consisting of members with a
common interest is found, advertising or promoting some products might be
very effective for members of the community due to the strong 
interest. 

Here, there are roughly two standpoints how we group communities. 
One is \textcolor{black}{context based} grouping, and the other is 
based on link structures. 
Previous work on community detection and grouping based on graph structure is summarized in \cite{Fortunato2010,Newman2010a,SCHAEFFER2007}, for example. 

Basically, these studies formulate network structure identification (community detection, grouping, and partition) as an optimization problem (sometimes it is not explicitly conscious), and design a fast algorithm to (approximately) solve the optimization problem. 
\textcolor{black}{Then, network structures to identify are obtained as outputs of the proposed algorithm. Here, network structures to identify are already abstract, e.g., dense subgraphs; the proposed algorithm can be used not only for the original purpose but also other purposes.}
In fact, \cite{Shi2000} is originally about boundary line detection in image data, but the proposed techniques are used for community detentions (e.g., \cite{Newman2010b}), and it is further used  
for the detection of industrial clusters in economic networks~\cite{Kagawa2013}. 
As above, the versatility of ``optimization problems'' is very useful.
However, we may think that they do not best utilize the features or characteristics 
of the target network. 
For example, the criteria for group partition, image processing, and detecting industrial clusters in economic networks could be different. In other words, we might expect a better performance by considering an optimization problem specialized for community detection. 

From these, we consider the problem for group partition (or simply say partition) in networks (graphs), taking into account the characteristics of SNS. In SNS, people communicate and exchange information with also a person who is not directly acquainted, i.e., followers. That is, in SNS, not only members with direct connections but also members without direct connections are loosely connected, which enables us to share information widely. Here, ``looseness'' is related to the degree of sharing information, and it is natural to define it as the distance (i.e., the length of a shortest path) between the persons on the network. 

Based on such observation, Branzei et al. introduced a new grouping scale for human relations networks~\cite{Branzei2011}. The definition of the utility in \cite{Branzei2011} is as follows: given a partition, the utility of an individual is defined as the sum of reciprocal distances to other people in the same coalition divided by the size of the coalition. Based on this, the social welfare of a partition is also defined as the sum of the utilities of all the members.
Unfortunately, finding a partition with maximum social welfare ({\sf MaxSWP}) is known to be NP-hard {even on graphs with maximum degree $6$~\cite{Balliu2017b}. 

Also, the characterizations of optimal partitions are known only for trivial cases such as complete graphs and complete bipartite graphs~\cite{Branzei2011}.
Even for trees, it is not known whether {\sf MaxSWP} can be solved in polynomial time.
One of the reasons seems to be the objective function of {\sf MaxSWP}. 
In a typical graph optimization problem, the objective function often forms a linear sum of weights, whereas 
the one of {\sf MaxSWP} takes the form of a nonlinear function, 
which is the sum of the reciprocal distances. 

\subsection{Our contribution} 
\color{black}
In this paper, we mainly study finding an optimal partition with social distance of a tree, which is 
 one of the most basic and important structures in graph algorithm design. 
In the process of research, we first give a complete characterization of optimal partitions of paths. 
Although the argument is simple, it gives an insight about the hardness related to the nonlinearity of the utility and the social welfare. 
Next, we give a similar characterization of optimal partitions of trees.
In the characterization, we  find out sub-trees with small diameters appeared in optimal partitions of trees. 
By using the characterization, we design a linear-time algorithm for computing an optimal partition of a tree. 
Finally, we show that {\sf MaxSWP} is NP-hard even for 4-regular graphs. This result strengthens the previous work for graphs with maximum degree $6$~\cite{Balliu2017b}. 
\color{black}

\subsection{Related work}\label{sec:relate}

\color{black}
Graph partition is \textcolor{black}{one of} the most basic and important problem in computer science and there are many \textcolor{black}{studies} about graph partition in various contexts, such as image processing and
cluster analysis~\cite{Shi2000,Fortunato2010,Newman2010a,SCHAEFFER2007,Kagawa2013}.

Graph partition with social distance has been studied in the context of
coalition formation games~\cite{Branzei2011}. 
In coalition formation games, each player has the utility based on the preference for other players in the same coalition.
Intuitively, a player is happy if the utility is high, that is, there are many players he/she prefer in the same coalition.
In the field of coalition formation games, many researchers study about desirable coalition formations, namely, partitions, in terms of maximum social welfare, stability, and core~\cite{AZIZ2013316,RAHWAN2015139}. 
Furthermore, the price of anarchy (PoA) and the price of stability (PoS) are also well-studied for evaluating agents systems~\cite{Balliu2017a}.
The PoA or PoS are more related to this paper because they are defined as the maximum and minimum ratio between a Nash stable solution and the best solution, respectively.




In coalition formation games on graphs, there are many utility functions for agents.
For example, in \cite{AZIZ2013316,SLESS2018217}, the utility of an agent is defined as the sum of edge-weights between him/her and other agents in the same coalition.
The weight of an edge represents the strength of the relationship between agents.
In social distance games, the utility is defined as the harmonic function of the distance between agents.
This is based on the concept of the closeness centrality, which is  one of classical measures for network analysis~\cite{Branzei2011,Balliu2017b}. 
\textcolor{black}{As mentioned above}, finding the best partition, that is, a partition with maximum social welfare is NP-hard even on graphs with maximum degree $6$~\cite{Balliu2017b}. 
On the other hand, there is a 2-approximaiton algorithm for finding such a partition~\cite{Branzei2011}. 
\color{black}

\bigskip

The organization of this paper is as follows.
In Section \ref{sec:pre}, we give basic terminologies, notation, and \textcolor{black}{definitions}.
In Section \ref{sec:path}, we give a complete characterization of optimal partitions of paths.  
In Section \ref{sec:tree}, 
we propose a linear-time algorithm for {\sf MaxSWP} on trees. 
Finally, we show that {\sf MaxSWP}  is NP-hard even on $4$-regular graphs in Section \ref{sec:NPh}.
\color{black}
Due to the space limitation, we out the proofs of propositions lemmas and theorems marked with (*). 
The detailed profs can be found in Appendix.
\color{black}

\section{Preliminaries}\label{sec:pre}
\subsection{Terminologies}
\color{black}
We use standard terminologies on graph theory.
Let $G=(V(G), E(G))$ be a simple, connected, and undirected graph.
For simplicity, we may denote $V(G)$ and $E(G)$ by $V$ and $E$, respectively.
We also denote the number of vertices and edges by $n$ and $m$, respectively.
A path from $u$ to $v$ of minimum length is called a {\em shortest path}, and the length is denoted by $dist_G(u,v)$. 
In a graph $G$, if there is no path from the vertex $u$ to the vertex $v$, we define $dist_G (u,v) =\infty$.
\color{black}
Let $G[C]$ be the subgraph induced by vertex set $C\subseteq{V}$. We sometimes denote $dist_{G[C]}(u,v)$ of $u,v \in{C}$ in $G[C]$  by $dist_{C}(u,v)$ for simplicity.
For graph $G$, we denote the {\em diameter} of $G$ by $diam(G)=\max_{u,v\in{V}, u\neq{v}}{dist_G (u, v)}$.
For a vertex $v$, we denote the set of neighbors of $v$ by $N(v)= \{u \mid (u,v)\in{E} \} \subseteq{V}$.
We also define the degree of $v$ as $d(v)=|{N(v)}|$.
For $G=(V,E)$, we denote the maximum degree of $G$ by $\Delta(G)=\max_{v\in V}d(v)$.
For simplicity, we sometimes denote it by $\Delta$.
For a positive integer $n$, we define $[n]=\{1, \ldots, n\}$.

A graph $G=(V,E)$ is called a {\em path graph} denoted by $P_n$ if $E=\{(v_i, v_{i +1})\mid{1} \le i<n\}$. We also sometimes simply call it a path.
Moreover, if $E=\{(v_1, v_{i})\mid{2} \le i\le n \}$, $G$ is said to be a {\em star} and denoted by $K_{1,n-1}$.
A graph $G$ is a {\em tree} if $G$ is connected and it has no cycle. We denote an $n$-vertex tree by $T_{n}$.
Moreover, we denote a tree with the diameter $d$ by $T^d_n$.
\color{black}

\subsection{Coalition and Utility}
\color{black}
The definitions here are based on~\cite{Branzei2011}.
\color{black}
Given a graph $G=(V,E)$ and $C\subseteq{V}$, we define the utility $U(v, C)$ of a vertex $v\in{C}$ as follows:
$$U(v,C)=\frac{1}{|{C}|}\sum_{u\in{C\backslash\{v\}}}\frac{1}{dist_{G[C]}(v, u)}.$$
By the definition, it satisfies that $0\leq U(v,C)\leq1$.
In a graph $G=(V,E)$, a {\em partition} of $G$ is defined as the family of sets of vertices ${\mathcal C}=\{C_1,\ldots, C_k\}$, where $C_1 \cup \cdots \cup C_k =V$ and $C_i \cap{C_j}=\emptyset$ for $i\neq{j}\in[k]$. 
Moreover, $C\in{\mathcal C}$ is called a {\em coalition} of partition ${\mathcal C}$.
In particular, if ${\mathcal C} = \{V\}$, ${\mathcal C}$ is called the {\em grand} of $G$ and $V$ is called the {\em grand coalition}.
If $\{v\} \in {\mathcal C}$ for a vertex $v\in V$, $v$ is said to be an {\em isolated vertex} of partition ${\mathcal C}$.
We define the utility of an isolated vertex as $U(v,\{v\})=0$.
Next, we define the {\em social welfare} of a partition ${\mathcal C}$ in graph $G$ as follows.
We define the {\em social welfare} $\varphi(G,{\mathcal C})$ of partition ${\mathcal C}$ in $G=(V,E)$ as follows:
$$\varphi(G,{\mathcal C})=\sum_{C\in {\mathcal C}} \sum_{v\in {C}}U(v, C).$$
If $\mathcal{C}$ is the grand of $G$, that is, $\mathcal{C}=\{V\}$, we simply denote $\varphi(G,\{V\})$ by $\varphi(G)$.
\textcolor{black}{We can observe that $\varphi(G,\mathcal{C})$ is bounded by $n-1$}.
Moreover, we define the {\em average social welfare} $\tilde{\varphi}(G, \mathcal{C})$ for partition $\mathcal{C}$ in $G$.
The {\em average social welfare} of partition $\mathcal{C}$ in $G=(V, E)$ is defined as follows:
$$\tilde{\varphi}(G,{\mathcal C})=\frac{\varphi(G,{\mathcal C})}{|{V}|}.$$
If $\mathcal{C}$ is the grand of $G$, that is, $\mathcal{C}=\{V\}$, we simply denote $\tilde{\varphi}(G,\{V\})$ by $\tilde{\varphi}(G)$.
Finally, we define a partition ${\mathcal C}^*$ with {\em maximum}  social welfare in graph $G$.
A partition $\mathcal{C^*}$ is {\em maximum} if it satisfies that $\varphi(G,{\mathcal C}^*) \ge \varphi(G,\mathcal{C})$ for any partition $\mathcal{C}$ in $G$.
\color{black}
We call the problem of finding a partition with maximum social welfare {\sf MaxSWP}.
We also call an optimal solution of {\sf MaxSWP} an {\em optimal partition}.
In previous work, it is shown that {\sf MaxSWP} is NP-hard even for graphs with maximum degree $6$~\cite{Balliu2017b}.
\color{black}
On the other hand, it is known that the grand is the only optimal partition of {\sf MaxSWP} on complete graphs and complete bipartite graphs~\cite{Branzei2011}.
\begin{proposition}[\cite{Branzei2011}]\label{opt:perfect}\rm
On complete graphs and complete bipartite graphs, the grand is the only optimal partition of {\sf MaxSWP}.
\end{proposition} 
\color{black}
Branzei et al. showed that there exists a partition where the utility of each vertex $v$ attains at least $1/2$ and a polynomial-time algorithm that finds such a partition for any graph~\cite{Branzei2011}. 
\begin{proposition}[\cite{Branzei2011}]\label{dia2}
\textcolor{black}{There is a polynomial-time algorithm that finds a partition such that each agent utility is at least $1/2$ for any graph}.
\end{proposition}
From Proposition~\ref{dia2}, it can be easily seen that there exists a partition $\mathcal{C}$ that satisfies $\varphi(G,{\mathcal C}) \ge{n/2}$. Thus, the social welfare of an optimal partition for any graph is also at least $n/2$.
\color{black}
\begin{corollary}\label{opt:partition}
Any optimal partition $\mathcal{C}^*$ of graph $G$ satisfies $\varphi(G,\mathcal{C}^*)\ge{n/2}$.
\end{corollary}
\color{black}
Since for any $G$ and $\mathcal{C}$, $\varphi(G,\mathcal{C})$ is bounded by $n-1$, the algorithm proposed by Branzei et al.~\cite{Branzei2011} is a $2$-approximation algorithm.
In the end of this section, we give another property of an optimal partition of {\sf MaxSWP}.
\begin{proposition}[{\bf *}]\label{connect:coalition}
For each coalition $C\in{\mathcal C}^*$ of optimal partition ${\mathcal C}^*$, $G[C]$ is connected.
\end{proposition}


\section{Optimal partition of a path}\label{sec:path}
In this section, we characterize the optimal partition of a path $P_n$.
In a path, the subgraph induced by a coalition is also a path by \textcolor{black}{Proposition~\ref{connect:coalition}}. 
By using this property and examining the average social welfare of $P_n$, we can identify the graph structures of coalitions in the optimal partitions of $P_n$.
In the following, we first examine average social welfare of $P_n$. 
Then we give the optimal partition of $P_n$.

Let $h(k)=\sum_{i = 1}^{k}1/i$ be the harmonic function for some positive integer $k$. The social welfare and the average social welfare of $P_n$ can be denoted by $\varphi(P_n)=(2\sum_{k=1}^{n-1}h(k))/n$ and $\tilde{\varphi}(P_n)=(2\sum_{k=1}^{n-1}h(k))/n^2$, respectively. 
Then, we obtain the following lemmas.
\begin{lemma}[{\bf *}]\label{lem:pathaverage}
It holds that $\tilde{\varphi}(P_2)<\tilde{\varphi}(P_3)$, and $\tilde{\varphi}(P_n)>\tilde{\varphi}(P_{n+1})$ for $n\ge3$.
\end{lemma}

\color{black}
\begin{lemma}[{\bf *}]\label{path:isolated}
For an optimal partition $\mathcal{C}^*$ of a path $P_n$ and a coalition $C\in\mathcal{C}^*$, $G[C]$ is either $P_2$, $P_3$ or $P_4$.
\end{lemma}
\color{black}

Finally, we give the optimal partition of $P_n$.
\begin{theorem}[{\bf *}]\label{th:pathopt}
The optimal partition of path $P_n$ is
\color{black}
\begin{enumerate}
\item $\mathcal{C}^*=\bigl\{\{v_{3i-2},v_{3i-1},v_{3i}\}\mid{1}\le{i}\le{n/3}\bigr\}$ if $n\equiv0\pmod{3}$,
\item $\mathcal{C}^*=\bigl\{\{v_1,v_2,v_3,v_4\},\{v_{3i+2},v_{3i+3},v_{3i+4}\}\mid{1}\le{i}\le{(n-4)/3}\bigr\}$ if $n\equiv1\pmod{3}$, and
\item $\mathcal{C}^*=\bigl\{\{v_1,v_2\},\{v_{3i},v_{3i+1},v_{3i+2}\}\mid{1}\le{i}\le{(n-2)/3}\bigr\}$ or $\bigl\{\{v_1,v_2,v_3,v_4\}, \break \{v_5,v_6,v_7,v_8\},\{v_{3i+6},v_{3i+7},v_{3i+8}\}\mid1\le{i}\le{(n-8)/3}\bigr\}$ if $n\equiv2\pmod{3}$.
\end{enumerate}
\color{black}
\end{theorem}

\section{Optimal partition of a tree}\label{sec:tree}
%
In Section \ref{sec:path}, we identified the optimal partition of {\sf MaxSWP} on a path.
In this section, we consider {\sf MaxSWP} on trees.
Since a tree is more general and complicated than a path and the optimal structure of  {\sf MaxSWP} is quite different from typical graph optimization problems, {\sf MaxSWP}  is non-trivial even on trees.

To solve {\sf MaxSWP},  we design an algorithm based on dynamic programming.
However, we do not know which information we keep track of in dynamic programming since the optimal structure of  {\sf MaxSWP} is unknown.
For this, we identify the small coalitions in the optimal partition. 
According to Corollary~\ref{opt:partition}, if there is a coalition whose social welfare is less than $n/2$, it is not included in the optimal partition since the social welfare can be increased by  dividing the coalition. 
Thus, we only keep track of the subgraph structures of coalitions of social welfare at least $n/2$. 
We can identify such coalitions by calculating the social welfare of each coalition. By using the subgraph structures of such coalitions,  we design a linear-time algorithm for {\sf MaxSWP} on a tree based on dynamic programming.

\subsection{Social welfare of trees with small diameters}\label{sec:opt-smalltree}
Since the utility of an agent is defined as the harmonic function with respect to the distance to all others in the same coalition, the diameter of the subgraph induced by each coalition affects the social welfare.
Intuitively, the social welfare of a subgraph with large diameter is very low in a tree because a tree is quite sparse. 
Therefore, we characterize the subgraph structures of coalitions in the optimal partition in terms of small diameters. 

We first consider trees $T^{\le2}_n$  with diameter at most 2.
Such graphs are stars denoted by $K_{1,n-1}$. 
Since a star is a complete bipartite graph, it satisfies that $\mathcal{C}^*=\{V\}$ by Property~\ref{opt:perfect}.
Here, we investigate the number of vertices of a star that maximizes the average social welfare. This gives the upper bound of the social welfare of $T^{\le2}_n=K_{1,n-1}$.

\begin{lemma}[{\bf *}]\label{max:ASW:diameter2}
For tree $T^{\le2}_n$ with diameter at most 2,  $n/2 \le \varphi(T^{\le2}_n) \le 9n/16$ holds. 
\end{lemma}

\begin{figure}[t] 
\begin{tabular}{cc}
\begin{minipage}{0.45\hsize}
\begin{center}
\includegraphics[width=35mm,keepaspectratio,clip]{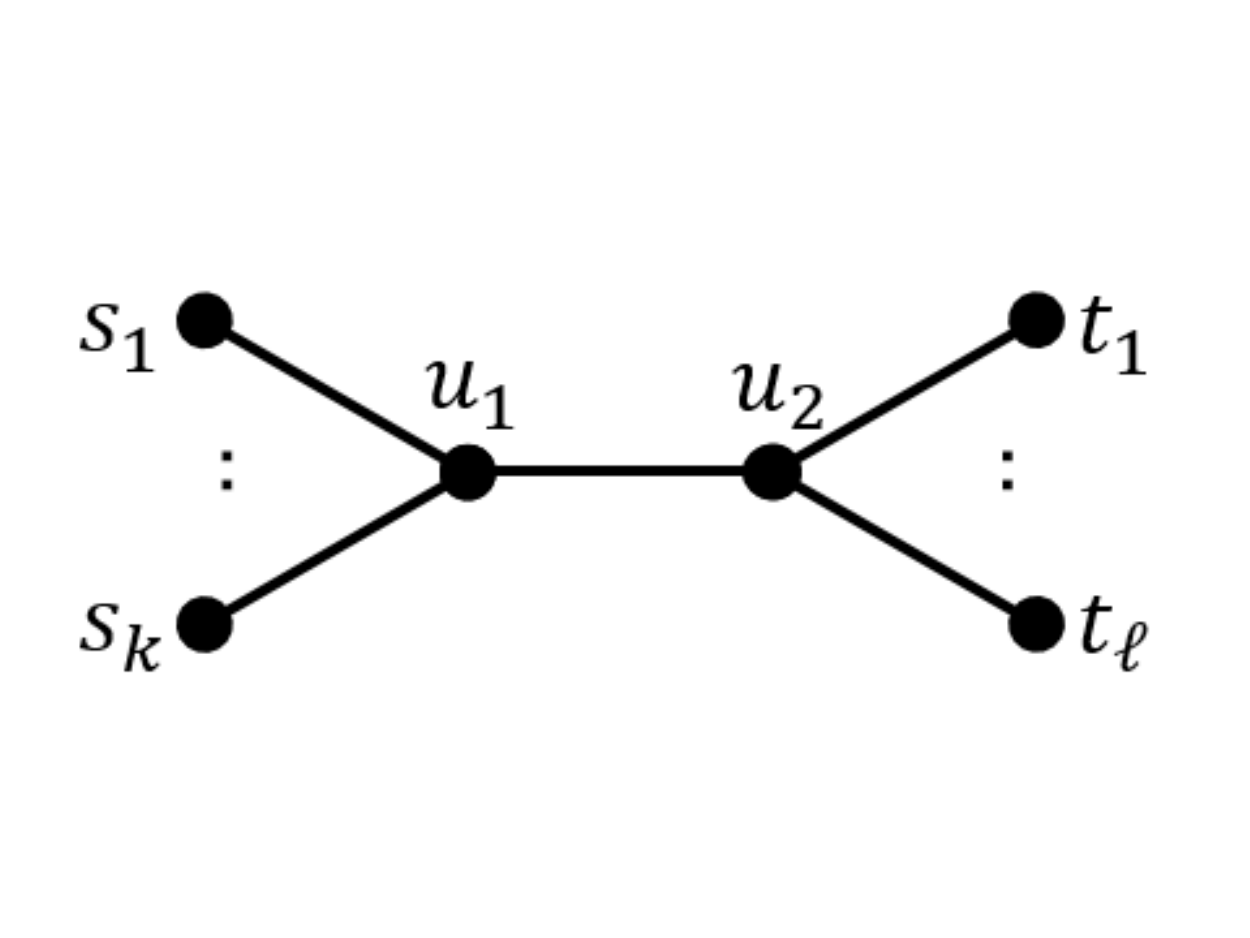}
\caption{diameter $3$ tree $T^3_n$}
\label{fig:diameter3tree}
\end{center}
\end{minipage}
\hspace*{.5cm}
\begin{minipage}{0.45\hsize}
\begin{center}
\includegraphics[width=35mm,keepaspectratio,clip]{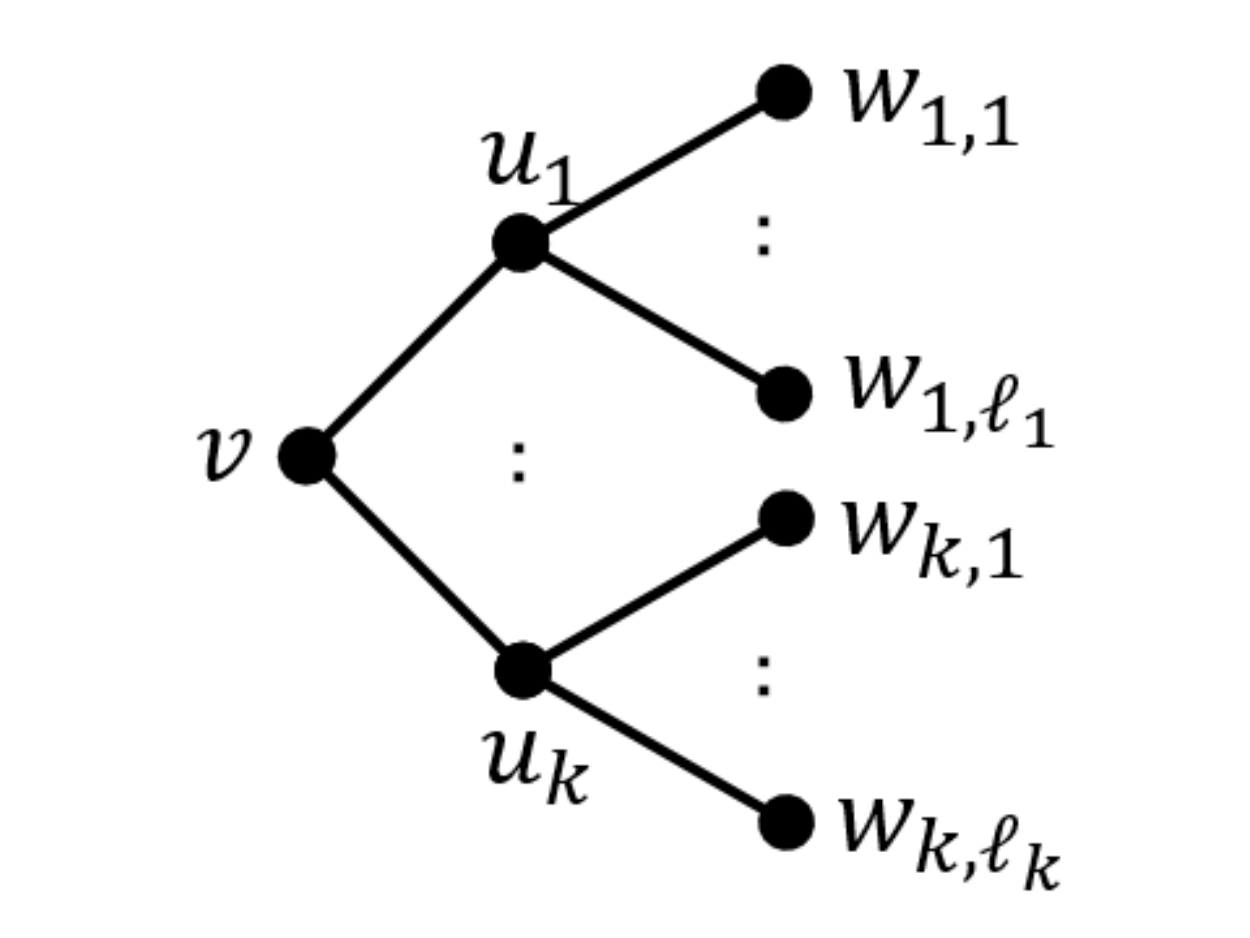}
\caption{diameter $4$ tree $T^4_n$}
\label{fig:diameter4tree}
\end{center}
\end{minipage}
\end{tabular}
\end{figure}

Next, we consider trees with diameter $3$. 
Any tree with diameter $3$ can be represented as $T^3_n=(V(T^3_n), E(T^3_n))$ where $V(T^3_n)=\{u_1,u_2,s_1,\ldots,s_k,t_1,\ldots, t_{\ell}\}$ and $E(T^3_n)=(u_1,u_2)\cup \{(u_1, s_i) \mid 1\le i\le k\}\cup \{(u_2, t_j) \mid 1\le j\le \ell\}$ for any $k,\ell \ge1$ (see Figure~\ref{fig:diameter3tree}).
Note that $n=|V(T^3_n)|=k+\ell+2$.
Then, the following lemma holds.
\begin{lemma}[{\bf *}]\label{SW:diameter3tree}
For tree $T^3_n$ with diameter 3, if $k=2$ and $\ell\ge 7$, $k \ge 7$ and $\ell=2$, $k>3$ and $\ell\ge 3$, or $k\ge 3$ and $\ell>3$, $\varphi(T^3_n)<n/2$ holds, and otherwise $\varphi(T^3_n)\ge n/2$.
\end{lemma}

%
%

As with trees with diameter 3, we identify the types of trees with diameter 4 that satisfy $\varphi(T^4_n)\ge n/2$.
Any tree with diameter 4 can be represented as $T^4_n=(V(T^4_n), E(T^4_n))$ where $V(T^4_n)=\{v, u_1,\ldots,u_k,w_{1,1},\ldots,w_{1, \ell_1},\ldots, w_{k,1},\ldots, w_{k, \ell_k}\}$ for $k\ge 2$ and $\ell_1, \ell_2\ge 1$, and $E(T^4_n)=\{(v, u_i)\mid 1\le i\le k\}\cup \{(u_i, w_{i, j})\mid 1\le i\le k, 1\le j \le \ell_k \}$ (see Figure~\ref{fig:diameter4tree}).
For each $i$, $\ell_i$ represents the number of leaves of $u_i$.
We denote the total number of leaves of $T^4_n$ by $\alpha_k=\sum_{i=1}^k\ell_i$ and then  the number of vertices of $T^4_n$ is represented as $n = |V(T^4_n)|=k+\alpha_k+1$.
Then, we obtain the following lemma.
\begin{lemma}[{\bf *}]\label{tree:diameter4}
For tree $T^4_n$ with diameter 4, $\varphi(T^4_n)\ge {n}/{2}$ holds if $(k,\alpha_k)=(2,2),(2,3),(3,2),(4,2)$, and otherwise $\varphi(T^4_n)<{n}/{2}$ holds.
\end{lemma}

Finally, we show that the social welfare of the grand coalition of a tree with diameter at least 5 is less than $n/2$.
\begin{lemma}[{\bf *}]\label{tree:diameter6}
For tree $T^\mu_n$ with diameter $\mu\ge5$, $\varphi(T^\mu_n)<n/2$ holds.
\end{lemma}

By the above discussion, the optimal partition of a tree does not contain not only coalitions with large diameters but also particular coalitions with small diameters.
In the following, we further refine the candidates for coalitions in the optimal partition of a tree.

First, we show that any optimal partition does not contain a coalition that consists of a tree with diameter 4. Thus, there is no coalition that consists of a tree with diameter at least 4 in the optimal solution by Lemma~\ref{tree:diameter6}.

\begin{lemma}[{\bf *}]\label{opt:diameter4tree}
Any optimal partition of a tree $T$ does not contain a coalition that consists of a tree $T^{\ge4}$ with diameter at least 4.
\end{lemma}

Moreover, we prove that the candidates for coalitions in the optimal partition are only three types.
\begin{lemma}[{\bf *}]\label{lem:treemaxSW}
Let $\mathcal{C}^*$ be the optimal partition of tree $T_n$.
Then, the subgraph $G[C]$ induced by $C\in{\mathcal C}^*$ is one of the following:
\color{black}
$(1)$ a star $K_{1,|C|-1}$ (see Figure~\ref{fig:star}), $(2)$ a path $P_4$ of length 3 (see Figure~\ref{fig:P_4}), and $(3)$ a tree $T^3_5$ of size five with diameter 3 (see Figure~\ref{fig:P'_4}).
\end{lemma}
\color{black}
\begin{figure}[t]
\begin{tabular}{cc}
\begin{minipage}{0.3\hsize}
\centering
\includegraphics[width=26mm,keepaspectratio,clip]{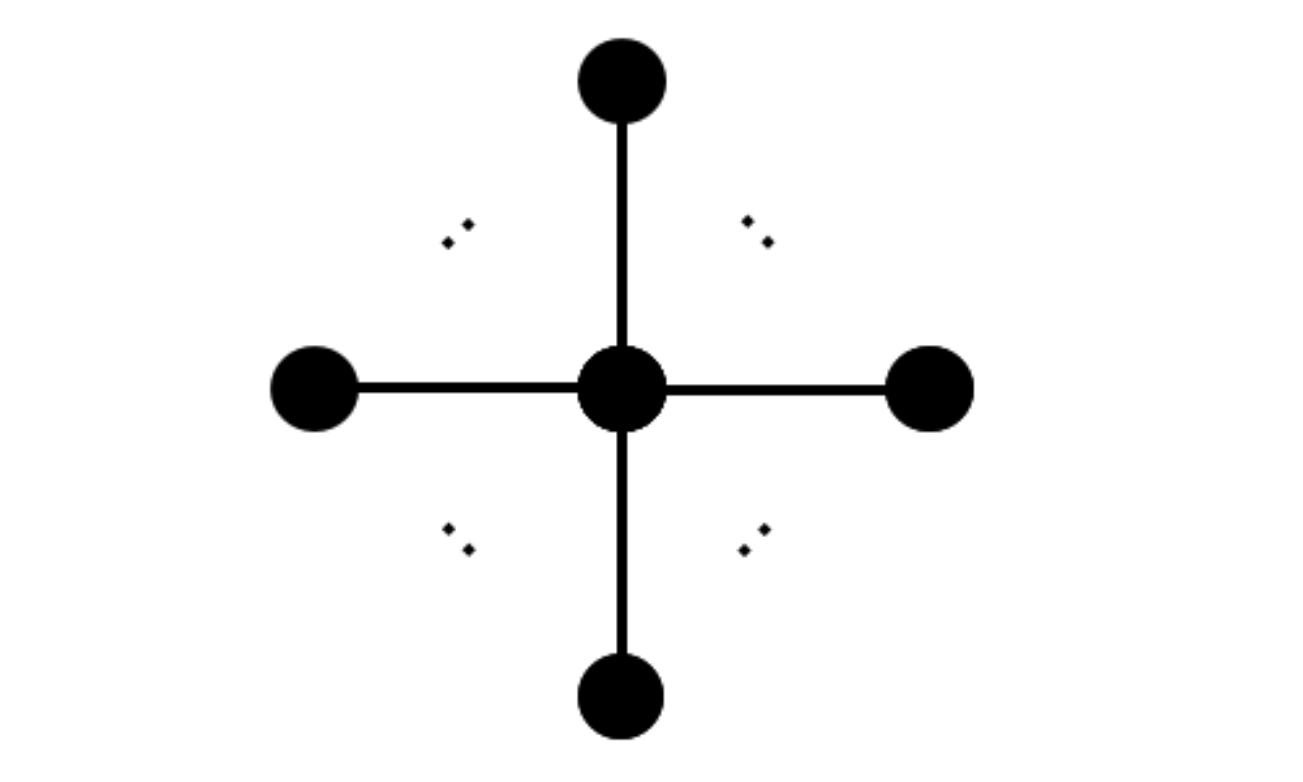}
\caption{$K_{1,|C|-1}$}
\label{fig:star}
\end{minipage}
\begin{minipage}{0.29\hsize}
\centering
\includegraphics[width=26mm,keepaspectratio,clip]{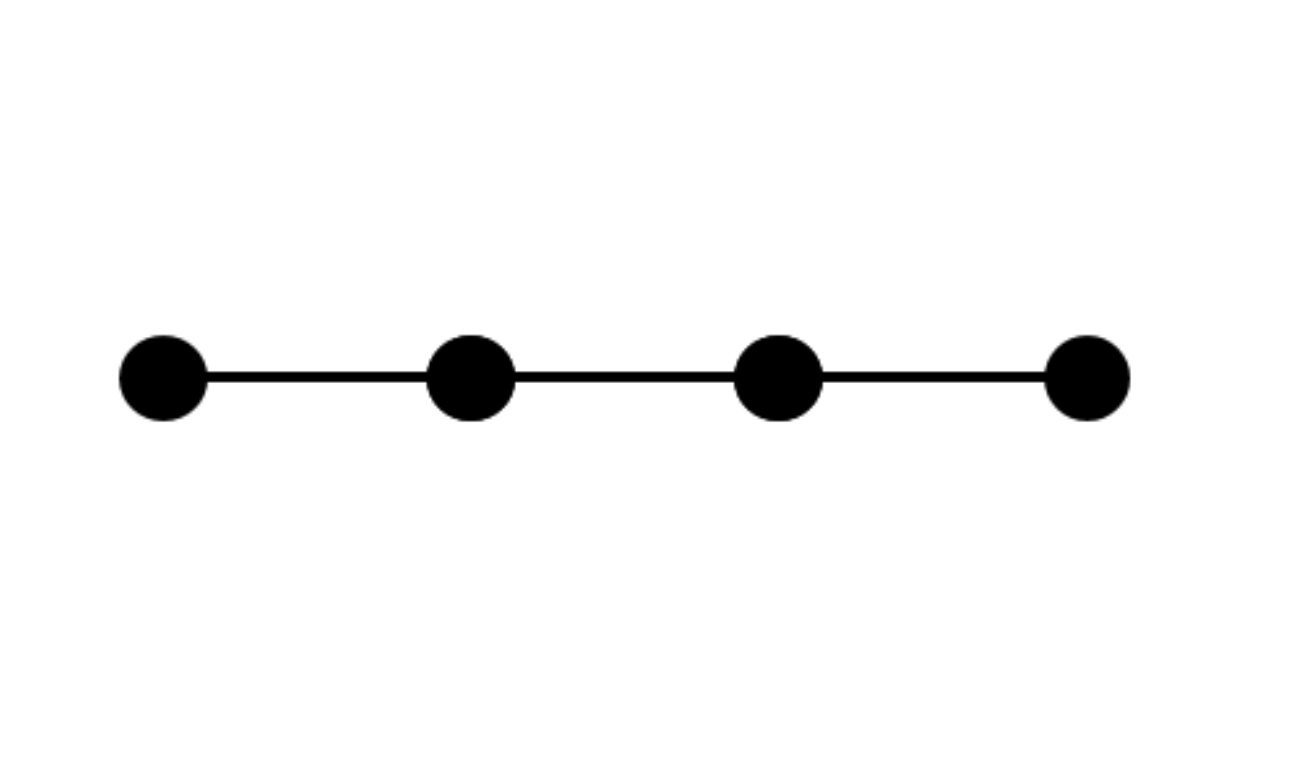}
\caption{$P_4$}
\label{fig:P_4}
\end{minipage}
\begin{minipage}{0.3\hsize}
\centering
\includegraphics[width=26mm,keepaspectratio,clip]{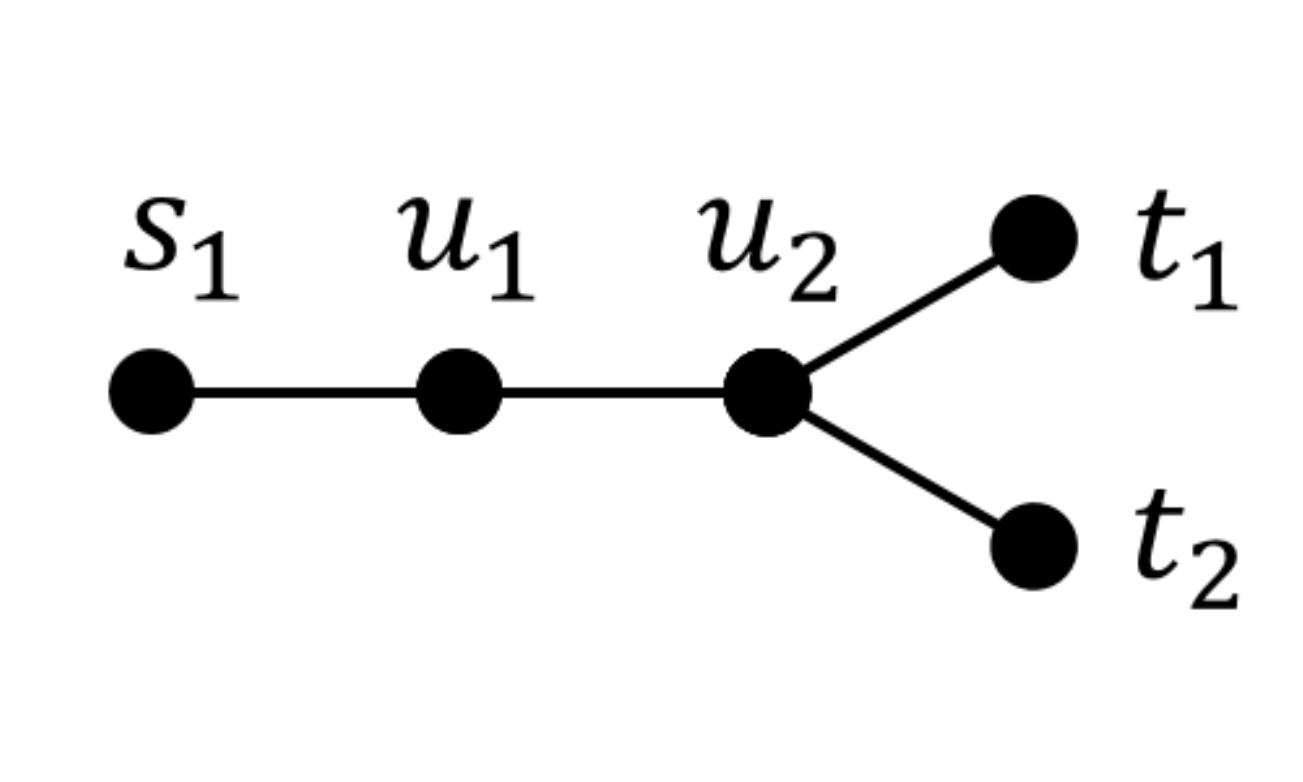}
\caption{$T^3_5$}
\label{fig:P'_4}
\end{minipage}
\end{tabular}
\end{figure}

\subsection{Algorithm}\label{sec:algo-opttree}
In this section, we propose an algorithm that finds an optimal partition of a tree in linear time.
This algorithm is based on dynamic programing with keeping track of the candidates identified by Lemma~\ref{lem:treemaxSW}.

First, we introduce some notations to design our algorithm.
Given a tree, we root it at arbitrary vertex $r$.
We denote a subtree whose root is $v\in{V}$ by $T_v$ and its partition by $\mathcal{C}_v$. 
For subtree $T_v$, we also denote the coalition including $v$ by $C_v\in \mathcal{C}_v$.

By Lemma~\ref{lem:treemaxSW}, the subgraph $G[C]$ induced by coalition $C\in {\mathcal C}^*$ is $K_{1, |C|-1}$, $P_4$ or $T^3_5$.
The algorithm recursively computes a partition of $T_v$ which attains the maximum social welfare for each $v$ from the leaves of $T$. 

Intuitively, our algorithm constructs coalitions in each step of dynamic programming. For example, a vertex $u$ is added a coalition as an isolated vertex in $T_u$. In next step, vertex $v$ must be added to the same coalition in  $T_v$ since the optimal solution does not contain an isolated vertex. 
Here, we keep track of not only coalition $C_v$, but also the position of $v$ in the coalition since we compute a coalition with maximum social welfare by combining sub-coalitions in subtrees of $T_v$.
For example, if $v$ is positioned at a leaf of $K_{1, f}$, it is combined with a coalition that consists of $K_{1,f-1}$. On the other hand, if $v$ is positioned at the center vertex of $K_{1, f}$,  it is combined with $f$ coalitions that consist of isolated vertices.
Then we compute each coalition and sub-coalition with maximum social welfare including a new vertex in next step again.
Since  the optimal partition contains only coalitions of $K_{1, |C|-1}$, $P_4$ and $T^3_5$, we keep track of coalition $C_v$ with the position of $v$ that consists of them.






Let $H$ be a subgraph induced by a coalition with the position of $v$.
Then we consider the following types of $H$:
\begin{enumerate} 
\item $H$ is an isolated vertex of $v$, denoted by $H=(\{v\},\emptyset)$, 

\item $H$ is a star $K_{1,f}$ and 
\begin{enumerate} 
\item  $v$ is the center of $K_{1,f}$, denoted by $H=K^{mid}_{1,f}$,
\item  $v$ is a leaf of $K_{1,f}$, denoted by $H=K^{leaf}_{1,f}$,
\end{enumerate}


\item $H$ is $P_4$ and 
\begin{enumerate} 
\item  $v$ is a leaf of $P_4$, denoted $H=P^{leaf}_4$,
\item  root $v$ is not a leaf of $P_4$, denoted by $H=P^{mid}_4$,
\end{enumerate}
\item $H$ is $T^3_5$ and
\begin{enumerate} 
\item  $v$ is $s_1$, denoted by $H=T^3_5(s_1)$,
\item  $v$ is $t_1$ or $t_2$, denoted by $H=T^3_5(t)$,
\item  $v$ is $u_1$, denoted by $H=T^3_5(u_1)$,
\item  $v$ is $u_2$, denoted by $H=T^3_5(u_2)$.
\end{enumerate}
\end{enumerate}

We can observe that connected proper subgraphs of $T^3_5$ are subgraphs of $P_4$ and star $K_{1,3}$. Also connected proper subgraphs of $P_4$ are subgraphs of star $K_{1,2}$.  Thus, by only keeping track of stars, we can treat $P_4$ and so $T^3_5$ as seen later.

Let $\mathcal{G}_v$ be the set of above subgraphs with the position of $v$ in $T_v$.
For  $T_v$, we define the recursive formula $\rho(v,H)=\max_{\mathcal{C}_v\ni{C_v}:G[C_v]=H}\varphi(T_v,\mathcal{C}_v)$ 
as the maximum social welfare of the partition  of $T_v$ such that  the subgraph induced by coalition $C_v$ including $v$ is $H\in {\mathcal{G}_v}$.
We also define $\rho(v)=\max_{H\in\mathcal{G}_v}\rho(v,H)$ as the  social welfare of the optimal partition of $T_v$.  
Then, the social welfare of the optimal partition of  $T$ with root $r$ is denoted by $\rho(r)=\max_{H\in\mathcal{G}_r}\rho(r,H)$.
\color{black}
Let  $w_j$ be the children of $v$ where $1\le j\le d(v)-1$ in $T_v$.
Then, we define the recursive formulas of $\rho(v,H)$ for $H\in {\mathcal{G}_v}$  to compute $\rho(r)$ as follows.


\begin{enumerate}
\item {\bf $H$ is an isolated vertex $(\{v\},\emptyset)$}

If $H=(\{v\},\emptyset)$, $\varphi(H)=0$.
Since $v$ separates trees $T_{w_j}$, $\rho(v,(\{v\},\emptyset))$ is the sum of the social welfare of the optimal partition in $T_{w_j}$.
Thus,  $\rho(v,(\{v\},\emptyset))$ is defined as $\rho(v,(\{v\},\emptyset))=\sum_{j=1}^{d(v)-1}\rho(w_j).$
%
\item {\bf $H$ is a star $K_{1,f}$}
\begin{enumerate}
\item {\bf $H=K^{mid}_{1,f}$ with center $v$}

In this case, we include $f$ children of $v$ in coalition $C_v$.
Note that $f$ children are isolated vertices in subtrees of $T_v$ since $C_v$ forms  $K_{1,f}$ in $T_v$.
Let $\delta_j=\rho(w_j)-\rho(w_j,(\{w_j\},\emptyset))$. Then $\delta_j$ means the difference between the maximum social welfare in $T_{w_j}$ and the maximum social welfare of the partition such that $w_j$ is an isolated vertex in $T_{w_j}$.
In other words, $\delta_j$ is the cost to include $w_j$ in $C_v$.
Thus, choosing the smallest $f$ children  of  $\delta_j$ maximizes $\rho(v,{K}^{mid}_{1,f}(v))$ since it consists of the social welfare of $C_v$, $f$ optimal partitions of $T_{w_j}$ such that $w_j$ is an isolated vertex in $T_{w_j}$, and $d(v)-1-f$ optimal partitions of $T_{w_j}$.
Let $w_1, w_2, \ldots, w_f$ be such children, where the indices are sorted in ascending order.
Then, $\rho(v,{K}^{mid}_{1,f}(v))$ is defined as 
$\rho(v,{K}^{mid}_{1,f}(v))=\varphi(K_{1,f})+\sum_{j=1}^{f}\rho(w_j, (\{w_j\}, \emptyset))+\sum_{j=f+1}^{d(v)}\rho(w_j)$.
%
\item {\bf $H=K^{leaf}_{1,f}$ with leaf $v$}

Since $C_v$ forms a star $K_{1,f}$ and $v$ is a leaf of it in $T_v$, we include vertex $v$ in a coalition of $K_{1,f-1}$ with center $w_k$ that is a child of $v$ in a subtree $T_{k}$. 
Thus, we need to choose such child $w_k$  that maximize the social welfare of $T_v$.
In this case, the maximum social welfare of $T_v$ is the sum of the social welfare of the optimal partition of subtrees $T_{w_j}$ except for $T_{w_k}$, $\varphi(K_{1,f})$, and the social welfare of the partition of $T_{w_k}$ such that $w_k$ is the center of $K_{1,f-1}$ of coalition $C_{w_k}$ minus $\varphi(K_{1,f-1})$. 
%

Thus, $\rho(v,K^{leaf}_{1,f})$ is defined as follows:
{\small
\begin{flalign*}
\hspace{-1.0cm}\rho(v,K^{leaf}_{1,f})
 &=\max_{k\in [d(v)]} \bigl\{ \sum_{j\in[d(v)]\setminus \{k\}}\rho(w_j)+\varphi(K_{1,f})+\rho(w_k,K^{mid}_{1,f-1}(w_k))-\varphi(K_{1,f-1})\bigr\}\\
&= \max_{k\in [d(v)]} \bigl\{ \sum_{j=1}^{d(v)} \rho(w_j)-\rho(w_k)+\varphi(K_{1,f})+\rho(w_k,K^{mid}_{1,f-1})-\varphi(K_{1,f-1})\bigr\}\\
&=  \sum_{j=1}^{d(v)} \rho(w_j)+\varphi(K_{1,f})-\varphi(K_{1,f-1})+\max_{k\in [d(v)]} \bigl\{\rho(w_k,K^{mid}_{1,f-1})-\rho(w_k)\bigr\}.
\end{flalign*}}

\end{enumerate}

We define $\rho(v,H)$ for the rest of $H\in {\mathcal G}_v$ in the same way.

\item {\bf $H$ is a path $P_4$}
\begin{enumerate}
\item  {\bf $H=P^{leaf}_4$ with leaf $v$}

A path whose one of leaves is $v$ consists of one $P_3=K_{1,2}$ and $v$.
 Thus we choose one child of $v$ whose coalition is $K_{1,2}$ and  maximizes   $\rho(v,P^{leaf}_4)$.
 {\small
\begin{flalign*}
\hspace{-1.0cm}\rho(v,P^{leaf}_4)=&\varphi(P_4)-\varphi(P_3)+\sum_{j=1}^{d(v)}\rho(w_j)+\max_{k\in[d(v)]}\left\{\rho(w_k,K^{leaf}_{1,2})-\rho(w_k)\right\}.
\end{flalign*}
}

\item {\bf $H=P^{mid}_4$ with non-leaf $v$}

A path whose one of non-leaf vertices is $v$ consists of one $P_2=K_{1,1}$, $v$, and one isolated vertex.
 Thus we choose two children of $v$ such that each coalitions that includes them is $K_{1,1}$ and an isolated vertex, respectively, and they maximize   $\rho(v,P^{mid}_4)$.
  {\small
\begin{flalign*}
\hspace{-1.0cm}\rho(v,P^{mid}_4)=&\varphi(P_4)-\varphi(P_2)+\sum_{j=1}^{d(v)}\rho(w_j)\\
&+\max_{a,b\in[d(v)],a\neq{b}}\bigl\{\rho(w_a,K_{1,1}^{leaf})+\rho(w_b,(\{w_b\},\emptyset))-\rho(w_a)-\rho(w_b)\bigr\}.
\end{flalign*}
}
\end{enumerate}

\item {\bf $H$ is a tree $T^3_5$}
\begin{enumerate}

\item {\bf $H=T^3_5(s_1)$ with $v=s_1$}

Since $v$ is $s_1$ of $T^3_5$ in $T_v$, we combine $K^{leaf}_{1,3}$ whose leaf is a child $w_j$ of $v$ with $v$.
Thus, we choose such a child of $v$ that maximizes  $\rho(v,T^3_5(s_1))$.
Then, $\rho(v,T^3_5(s_1))$ is defined as follows:
{\small
\begin{flalign*}
\hspace{-1.0cm}\rho(v,T^3_5(s_1))=\varphi(T^3_5)-\varphi(K_{1,3})+\sum_{j=1}^{d(v)}\rho(w_j)+\max_{k\in[d(v)]}\left\{\rho(w_k,K^{leaf}_{1,3})-\rho(w_k)\right\}.
\end{flalign*}
}


\item {\bf $H=T^3_5(t)$ with $v=t_1$ or $t_2$}

Since $v$ is $t_1$ or $t_2$ of $T^3_5$ in $T_v$, we combine $P^{mid}_4$ in a subtree $T_{w_j}$ with $v$.
Thus, we choose such a child of $v$ that maximizes  $\rho(v,T^3_5(t))$.
Then, $\rho(v,T^3_5(t))$ is defined as follows:
{\small
\begin{flalign*}
\hspace{-1.0cm}\rho(v,T^3_5(t))=\varphi(T^3_5)-\varphi(P_4)+\sum_{j=1}^{d(v)}\rho(w_j)+\max_{k\in[d(v)]}\left\{\rho(w_k,P^{mid}_4)-\rho(w_k)\right\}.
\end{flalign*}
}

\item {\bf $H=T^3_5(u_1)$ with $v=u_1$}

Since $v$ is $u_1$ of $T^3_5$ in $T_v$, we combine one coalition of $K_{1,1}$ whose center is a child of $v$, one coalition of an isolated vertex, and $v$ to construct coalition $C_v$.
Thus, we choose two such children of $v$ that maximizes  $\rho(v,T^3_5(u_1))$.
Then, $\rho(v,T^3_5(u_1))$ is defined as follows:
{\small
\begin{flalign*}
\hspace{-1.0cm}\rho(v,T^3_5(u_1))=&\varphi(T^3_5)-\varphi(P_3)+\sum_{j=1}^{d(v)}\rho(w_j)\\
&+\max_{a,b\in[d(v)],a\neq{b}}\bigl\{\rho(w_a,K^{mid}_{1,2}(w_a))+\rho(w_b,(\{w_b\},\emptyset))-\rho(w_a)-\rho(w_b)\bigr\}.
\end{flalign*}
}
\item {\bf $H=T^3_5(u_2)$ with $v=u_2$}

Since $v$ is $u_2$ of $T^3_5$ in $T_v$, we combine one coalition of $P_2$ whose leaf is a child of $v$, two coalitions of isolated vertices, and $v$  to construct coalition $C_v$.
Note that such $P_2$ is a star $K^{leaf}_{1,1}$.
Thus, we choose three such children of $v$ that maximizes  $\rho(v,T^3_5(u_2))$.
Then $\rho(v,T^3_5(u_2))$ is defined as follows:
{\small
\begin{flalign*}
\hspace{-1.0cm}\rho(v,T^3_5(u_2))=&\varphi(T^3_5)-\varphi(K^{leaf}_{1,1})+\sum_{j=1}^{d(v)}\rho(w_j)\\
&+\max_{a,b,c\in[d(v)],a\neq{b}\neq{c}}\bigl\{\rho(w_a,K^{leaf}_{1,1})+\rho(w_b,(\{w_b\},\emptyset))+\rho(w_c,(\{w_c\},\emptyset))\\
&\hspace{3cm}-\rho(w_a)-\rho(w_b)-\rho(w_c)\bigr\}.
\end{flalign*}
}
\end{enumerate}
\end{enumerate}

\begin{figure}[t]
\centering
\includegraphics[width=70mm]{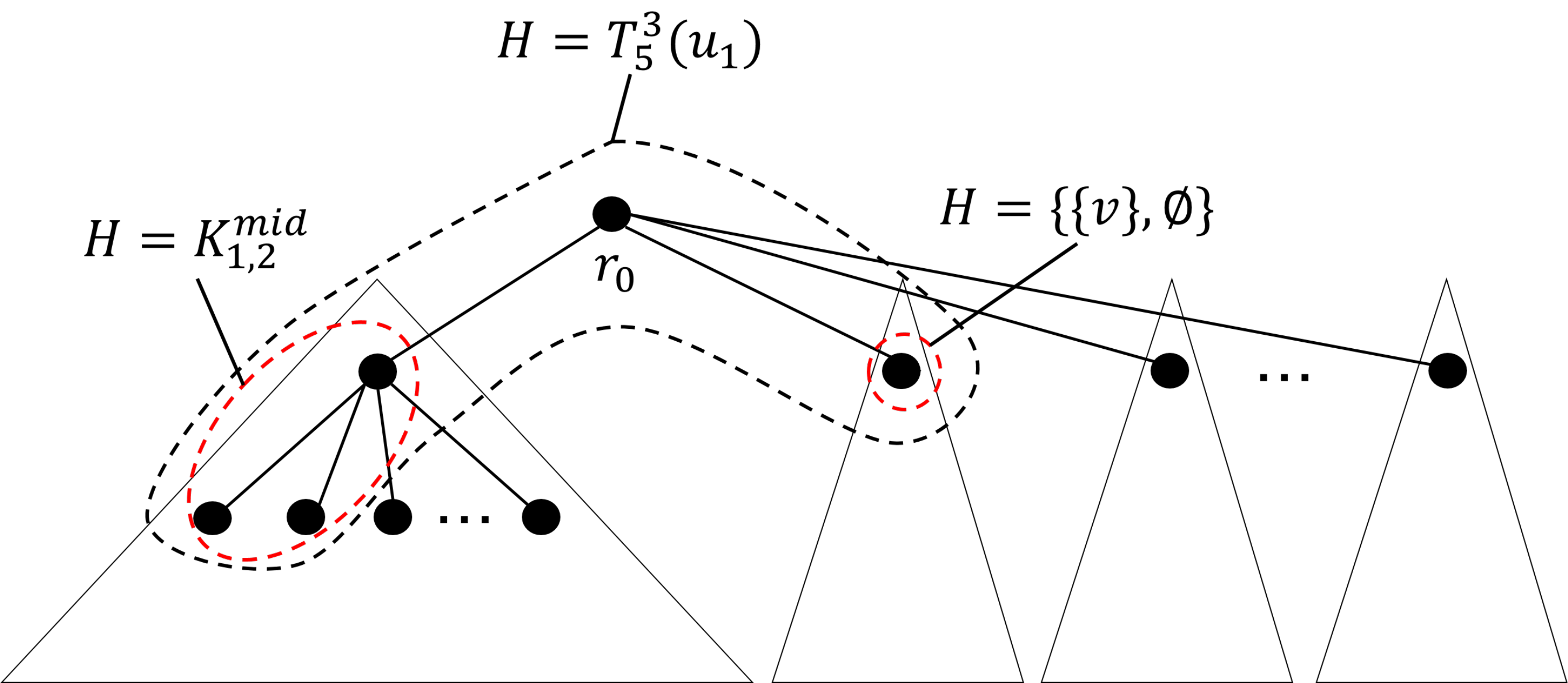}
\caption{Computing $\rho(r_0,H)$ for $H=T^3_5(u_1)$}
\label{fig:dpalgorithm3}
\end{figure}
Figure~\ref{fig:dpalgorithm3} shows an example of computing $\rho(r_0,H)$ where $H=T^3_5(u_1)$. 
To compute $\rho(r_0,H)$, we use the $\rho$'s values of its subtrees.
The pattern $H=T^3_5(u_1)$ contains one subtree with $H=K^{mid}_{1,2}$ and one with $H=\{\{v\},\emptyset \}$.
The best combination of these can be computed by the DP procedure $(c)$ explained above.

Finally, we evaluate the running time of our algorithm.
In Case 1, we can compute the recursive formula in time $O(d(v))$.
In Case 2, for case (a), we need to compute largest $\rho(v,{K}^{mid}_{1,f}(v))$ among $f=1,2,\ldots,d(v)-1$. This can be done by a binary search with SELECT, 
since $\delta_i$ is increasing and $w_i$'s utility in $K_{1,f}$ is decreasing. 
We can find the optimal $f$ in $O(d(v)+d(v)/2 + d(v)/4 \cdots)=O(d(v))$. 
Case (b) is also computable in the same running time. In Case 3, both cases can be computed in time $O(d(v))$ by memorizing the best score among its children. 
Finally, in Case 4, all the cases can be computed in $O(d(v))$ by a similar manner of Case 3. 
Thus the total running time of this algorithm is $\sum_{v\in{V}}{O(d(v))}=O(n)$, since $\sum_{v\in{V}}d(v)=2|E|=2(n-1)$ holds for a tree by the handshaking lemma.
Hence, we obtain the following theorem.
\begin{theorem}\label{th:treeoptDP}
 {\sf MaxSWP} for a tree can be solved in linear time.
\end{theorem}
\color{black}

\section{Hardness result of {\sf MaxSWP} for 4-regular graphs}~\label{sec:NPh}
\color{black}
It is mentioned in \cite{Balliu2017b} that {\sf MaxSWP} is NP-hard for graphs with maximum degree $6$, though the proof is omitted in the conference paper. 
Actually, we can show a stronger result, that is, {\sf MaxSWP} is NP-hard even for $4$-regular graphs. 
The proof is based on a reduction from a restricted variant of $3$-SAT problem called {\sf M3XSAT(3L)}, which is shown to be NP-complete in \cite[Lemma 5]{Porschen2014}. 
\color{black}

\begin{theorem}[{\bf *}]\label{nphard:4-regular}
{\sf MaxSWP} is NP-hard even for 4-regular graphs. 
\end{theorem}


\bibliographystyle{splncs04} 
\bibliography{reference.bib} 

\begin{thebibliography}{10}
\providecommand{\url}[1]{\texttt{#1}}
\providecommand{\urlprefix}{URL }
\providecommand{\doi}[1]{https://doi.org/#1}

\bibitem{AZIZ2013316}
Aziz, H., Brandt, F., Seedig, H.G.: Computing desirable partitions in
  additively separable hedonic games. Artificial Intelligence  \textbf{195},
  316 -- 334 (2013)

\bibitem{Balliu2017a}
Balliu, A., Flammini, M., Melideo, G., Olivetti, D.: Nash stability in social
  distance games. In: Proceedings of the Thirty-First {AAAI} Conference on
  Artificial Intelligence ({AAAI 2017}). pp. 342--348 (2017)

\bibitem{Balliu2017b}
Balliu, A., Flammini, M., Olivetti, D.: On pareto optimality in social distance
  games. In: Proceedings of the Thirty-First {AAAI} Conference on Artificial
  Intelligence ({AAAI} 2017). pp. 349--355 (2017)

\bibitem{Branzei2011}
Br{\^{a}}nzei, S., Larson, K.: Social distance games. In: Proceedings of the
  22nd International Joint Conference on Artificial Intelligence ({IJCAI}
  2011). pp. 91--96 (2011)

\bibitem{Fortunato2010}
Fortunato, S.: Community detection in graphs. Physics Reports  \textbf{486}(3),
   75--174 (2010)

\bibitem{Kagawa2013}
Kagawa, S., Okamoto, S., Suh, S., Kondo, Y., Nansai, K.: Finding
  environmentally important industry clusters: Multiway cut approach using
  nonnegative matrix factorization. Social Networks  \textbf{35}(3),  423--438
  (2013)

\bibitem{Newman2010a}
Newman, M.E.J.: Networks: An Introduction. Oxford University Press (2010)

\bibitem{Newman2010b}
Newman, M.E.J.: Spectral methods for community detection and graph
  partitioning. Physical Review E  \textbf{88},  042822 (2013)

\bibitem{Porschen2014}
Porschen, S., Schmidt, T., Speckenmeyer, E., Wotzlaw, A.: {XSAT and NAE-SAT of
  Linear CNF Classes}. Discrete Appl. Math.  \textbf{167},  1--14 (2014).
  \doi{10.1016/j.dam.2013.10.030}

\bibitem{RAHWAN2015139}
Rahwan, T., Michalak, T.P., Wooldridge, M., Jennings, N.R.: Coalition structure
  generation: A survey. Artificial Intelligence  \textbf{229},  139 -- 174
  (2015)

\bibitem{SCHAEFFER2007}
Schaeffer, S.E.: Graph clustering. Computer Science Review  \textbf{1}(1),
  27--64 (2007)

\bibitem{Shi2000}
Shi, J., Malik, J.: Normalized cuts and image segmentation. IEEE Transactions
  on Pattern Analysis and Machine Intelligence  \textbf{22}(8),  888--905
  (2000)

\bibitem{SLESS2018217}
Sless, L., Hazon, N., Kraus, S., Wooldridge, M.: Forming $k$ coalitions and
  facilitating relationships in social networks. Artificial Intelligence
  \textbf{259},  217 -- 245 (2018)

\end{thebibliography}
\section{Omitted proof in Section~\ref{sec:pre}}
\let\temp\theproposition
\renewcommand{\theproposition}{\ref{connect:coalition}}
\begin{proposition}
For each coalition $C\in{\mathcal C}^*$ of optimal partition ${\mathcal C}^*$, $G[C]$ is connected.
\end{proposition}
\let\theproposition\temp
\begin{proof}\rm
Let $\mathcal {C}^*$ be an optimal partition of $G$.
Suppose that there is a coalition $C\in \mathcal{C}^*$ such that $G[C]$ has at least two connected components. 
Note that $1\leq|C_1|, |C_2|<|C|$.
We also set $\mathcal{C}'=\mathcal{C}^* \setminus \{C\} \cup \{C_1,C_2\}$.
\textcolor{black}{By using the fact that $|C_1|<|C|$, for any $v\in C_1\cup C$} it satisfies that
\begin{flalign*}
U(v,C)=\frac{1}{|{C}|}\sum_{u\in{C\setminus \{v\}}}\frac{1}{dist_{G[C]}(v, u)}&=\frac{1}{|{C}|}\sum_{u\in{C_1\setminus \{v\}}}\frac{1}{dist_{G[C]}(v, u)}\\
&<\frac{1}{|{C_1}|}\sum_{u\in{C_1\setminus \{v\}}}\frac{1}{dist_{G[C_1]}(v, u)}=U(v,C_1).
\end{flalign*}
In the same way, we obtain $U(v,C)<U(v,C_2)$ for any $v\in C_2\cup C_1$.
Thus,
\begin{align*}
\sum_{v\in C} U(v,C)=\sum_{v\in C_1} U(v,C)+\sum_{v\in C_2} U(v,C)< \sum_{v\in C_1} U(v,C_1) +\sum_{v\in C_2} U(v,C_2).
\end{align*}
%
Let $\varphi(G,\mathcal{C^*})$ be the social welfare of $\mathcal{C^*}$ in $G$. Because the social welfare of $\mathcal{C'}$ in $G$ is $\varphi(G,{\mathcal C}')=\varphi(G,{\mathcal C}^*)-\sum_{v\in C}U(v,C)$ +\textcolor{black}{$\sum_{v\in C_1} U(v,C_1) +\sum_{v\in C_2} U(v,C_2)$} , it holds that $\varphi(G,{\mathcal C}')>\varphi(G,{\mathcal C}^*)$.
This contradicts that ${\mathcal C}^*$ is optimal. 
\qed \end{proof}

\section{Omitted proof in Section~\ref{sec:path}}

\let\temp\thelemma
\renewcommand{\thelemma}{\ref{lem:pathaverage}}
\begin{lemma}
It holds that $\tilde{\varphi}(P_2)<\tilde{\varphi}(P_3)$, and $\tilde{\varphi}(P_n)>\tilde{\varphi}(P_{n+1})$ for $n\ge3$.
\end{lemma}
\let\thelemma\temp
\addtocounter{lemma}{-1}
\begin{proof}\rm
First, we compute the average social welfare of $P_n$. 
\small
\color{black}
\begin{flalign*}
\tilde{\varphi}(P_n)=\frac{2}{n^2}\sum_{k=1}^{n-1}h(k)&=\frac{2}{n^2}\left(\frac{1}{n-1}+\frac{2}{n-2}+\cdots+\frac{n-1}{1}\right)\\
&=\frac{2}{n^2}\left(\frac{n-(n-1)}{n-1}+\frac{n-(n-2)}{n-2}+\cdots+\frac{n-1}{1}\right)\\
&=\frac{2}{n^2}\left(\frac{n}{n-1}+\frac{n}{n-2}+\cdots+\frac{n}{1}-1\cdot(n-1)\right)\\
&=\frac{2}{n^2}\left(n\cdot h(n-1)+1-n\right).
\end{flalign*}
\normalsize
Next, we consider the difference between $\tilde{\varphi}(P_{n+1})$ and $\tilde{\varphi}(P_n)$.
{\small
\begin{flalign*}
&\tilde{\varphi}(P_{n+1})-\tilde{\varphi}(P_n)\\
&=\frac{2}{n^2(n+1)^2}\bigl(n^2(n\cdot h(n)+h(n)-n)-(n+1)^2(n\cdot h(n-1)+1-n)\bigr)\\
&=\frac{2}{n^2(n+1)^2}\bigl((n^3\cdot h(n)+n^2\cdot h(n)-n^3)-(n^3\cdot h(n-1)+n^2-n^3)\\
&\hspace{5.0cm}-(2n+1)(n\cdot h(n-1)+1-n)\bigr)\\
&=\frac{2}{n^2(n+1)^2}\left(n^2\cdot h(n)-(2n+1)(n\cdot h(n-1)+1-n)\right)\\
&=\frac{2}{n^2(n+1)^2}\left(2n^2-n^2\cdot h(n-1)-n\cdot h(n-1)-1\right).
\end{flalign*}}

Let $q_n=n^2(2-h(n-1))-n\cdot h(n-1)-1$ for $n$. Then the sign of $q_n$ and the sign of $\tilde{\varphi}(P_{n+1})-\tilde{\varphi}(P_n)$ are the same.
\color{black}
We also note that $q_n$ is always negative for $n\ge3$ and positive for $n=2$. Thus, we obtain $\tilde{\varphi}(P_2)<\tilde{\varphi}(P_3)$, and $\tilde{\varphi}(P_n)>\tilde{\varphi}(P_{n+1})$ for $n\ge3$.
\qed \end{proof}


\let\temp\thelemma
\renewcommand{\thelemma}{\ref{path:isolated}}
\begin{lemma}
For the optimal partition $\mathcal{C}^*$ of a path $P_n$ and coalition $C\in\mathcal{C}^*$, $G[C]$ is either $P_2$, $P_3$ or $P_4$.
\end{lemma}
\let\thelemma\temp
\addtocounter{lemma}{-1}
\begin{proof}\rm
We first observe that the candidates for $G[C]$ are only $P_2, P_3, P_4, P_5 $ or an isolated vertex.
%
This is because we can increase the social welfare by dividing $P_n$ for $n>5$ to them from Corollary~\ref{opt:partition}, Lemma~\ref{lem:pathaverage}, and $\varphi(P_n)<n/2$ for $\textcolor{black}{n>5}$.

Next, we show that $P_5$ is not included in the candidate.
Since we have $\varphi(P_5)=77/30<8/3=\varphi(P_2)+\varphi(P_3)$, we can increase the social welfare by dividing $P_5$ into two coalition $V(P_2)$ and $V(P_3)$, if there is a coalition such that $G[C]=P_5$.
Therefore $P_5$ is not included in the candidate for $G[C]$.

Finally, we show that an isolated vertex is not the candidate.
Suppose that an isolated vertex is included in the optimal partition of $P_n$.
Because a graph is a path graph, any isolated vertex is adjacent to at least one vertex.
Now, we suppose that an isolated vertex is adjacent to a coalition that consists of another isolated vertex, $P_2$, $P_3$, or $P_4$.
In the following, we show the contradiction for these four cases.
\begin{enumerate} 
\item If an isolated vertex $v$ is adjacent to an isolated vertex $u$, we set $\varphi(P_n,\mathcal{C}^*\setminus \{\{u\},\{v\}\} \cup \{\{u,v\}\})=\varphi(P_n,\mathcal{C}^*)-0+1$. This contradicts the optimality of $\varphi(P_n,\mathcal{C}^*)$.
\item If an isolated vertex $v$ is adjacent to a vertex of the coalition $V (P_2)$, we set $\varphi(P_n,\mathcal{C}^*\setminus \{V(P_2),\{v\}\}\cup \{V(P_3\})=\varphi(P_n,\mathcal{C}^*)-1+5/3$. This contradicts the optimality of $\varphi(P_n,\mathcal{C}^*)$.
\item If an isolated vertex $v$ is adjacent to a vertex of the coalition $V(P_3)$, we set $\varphi(P_n,\mathcal{C}^*\setminus \{V(P_3),\{v\}\}\cup \{V(P_4\})=\varphi(P_n,\mathcal{C}^*)-5/3+13/6$. This contradicts the optimality of $\varphi(P_n,\mathcal{C}^*)$.

\item If an isolated vertex $v$ is adjacent to a vertex of the coalition $V(P_4)$, we set $\varphi(P_n,\mathcal{C}^*\setminus \{V(P_4),\{v\}\}\cup \{V(P_2),V(P_3)\})=\varphi(P_n,\mathcal{C}^*)-13/6+8/3$. This contradicts the optimality of $\varphi(P_n,\mathcal{C}^*)$.
\end{enumerate}
In any case, an isolated vertex is not included in the optimal partition.
Therefore, $G[C]$ is $P_2,P_3$ or $P_4$ in the optimal partition of $P_n$. 
\qed \end{proof}

\let\temp\thetheorem
\renewcommand{\thetheorem}{\ref{th:pathopt}}
\begin{theorem}
The following partitions are optimal partitions of path $P_n$:
\color{black}
\begin{enumerate}
\item $\mathcal{C}=\bigl\{\{v_{3i-2},v_{3i-1},v_{3i}\}\mid{1}\le{i}\le{n/3}\bigr\}$ if $n\equiv0\pmod{3}$,
\item $\mathcal{C}=\bigl\{\{v_1,v_2,v_3,v_4\},\{v_{3i+2},v_{3i+3},v_{3i+4}\}\mid{1}\le{i}\le{(n-4)/3}\bigr\}$ if $n\equiv1\pmod{3}$, and
\item $\mathcal{C}=\bigl\{\{v_1,v_2\},\{v_{3i},v_{3i+1},v_{3i+2}\}\mid{1}\le{i}\le{(n-2)/3}\bigr\}$ or $\bigl\{\{v_1,v_2,v_3,v_4\}, \break\{v_5,v_6,v_7,v_8\},\{v_{3i+6},v_{3i+7},v_{3i+8}\} \mid1\le{i}\le{(n-8)/3}\bigr\}$ if $n\equiv2\pmod{3}$.
\end{enumerate}
\color{black}
\end{theorem}
\let\thetheorem\temp
\addtocounter{theorem}{-1}
\begin{proof}\rm
From Lemma~\ref{path:isolated}, we only have to consider partitions that consist of $P_2$,$P_3$ or $P_4$.
Note that the order of coalitions in a path graph does not affect.
Thus, we can rearrange the order of coalitions in a path graph without changing the social welfare.
Now, we show that the optimal solution of $P_n$ contains at most two coalitions of $V(P_2)$.
If three or more coalitions of $V(P_2)$ are included in the optimal partition, we first rearrange three coalitions of $V(P_2)$ so that they are consecutive without changing the social welfare. Next, we replace them by consecutive two coalitions of $V(P_3)$. Because $\varphi(P_2)=1$, $\varphi(P_3)=5/3$ and $3\cdot \varphi(P_2)=3<10/3=2\cdot \varphi(P_3)$, the social welfare increases.
Similarly, we show that the optimal partiton of $P_n$ contains at most two coalitions of $V(P_4)$. If three or more coalitions of $V(P_4)$ are included in the optimal partition, we first rearrange three coalitions of $V(P_4)$ so that they are consecutive. Then, we change them to consecutive four coalitions of $V(P_3)$. Because $\varphi(P_4)=13/6$, $\varphi(P_3)=5/3$ and $3\cdot \varphi(P_4)=13/2<20/3=4\cdot \varphi(P_3)$, the social welfare also increases.
Thus, the optimal partition of $P_n$ contains at most two coalitions of $V(P_2)$ and $V(P_4)$, respectively.

Moreover, if both $V(P_2)$ and $V(P_4)$ are included in the optimal partition, we rearrange one $V(P_2)$ and one $V(P_4)$ so that they are consecutive. Then, we replace them by two consecutive consists of $V(P_3)$. Because $\varphi(P_2)+\varphi(P_4)=19/6<10/3=2\cdot \varphi(P_3)$, the social welfare increases.
Thus, both $V(P_2)$ and $V(P_4)$ are not included in the optimal partition.

Finally, if the optimal partiton of $P_n$ contains two coalitions of $V(P_2)$, we rearrange them so that so that they are consecutive.
Next, we change two $V(P_2)$ into one $V(P_4)$. Because $2\cdot \varphi(P_2)=2<13/6=\varphi(P_4)$, the social welfare increases.
Thus, the optimal partition of $P_n$ contains at most one coalition of $V(P_2)$.
Therefore, the optimal partition of $P_n$ contains at most one coalition of $V(P_2)$, at most two coalitions of $V(P_4)$. Moreover, it does not contain both $V(P_2)$ and $V(P_4)$.

From the above discussion, we can identify the optimal partition of a path graph $P_n$.
\begin{enumerate}
\item If $n\equiv0\pmod{3}$,\\
A partition $\mathcal{C}=\{V(P_3), \ldots, V(P_3) \}$ satisfies the above conditions.
That is, the partition that consists of coalitions $V(P_3)$'s is the optimal partition.
\item If $n\equiv1\pmod{3}$,\\
A partition $\mathcal{C}=\{V(P_4),V(P_3),\ldots,V(P_3)\}$ is the only partition that satisfies the above condition.
\item If $n\equiv2\pmod{3}$,\\
Partitions $\mathcal{C}=\{V(P_2),V(P_3),\ldots,V(P_3)\}$ and $\mathcal{C}=\{V(P_4), V(P_4), V(P_3),\break\ldots, V(P_3)\}$ satisfy the above conditions.
The social welfare of the former is $1+(5/3)\cdot(n-2)/3=(5n-1)/9$ and the latter is $(13/6)\cdot2+(5/3)\cdot(n-8)/3=(5n-1)/9$. 
Thus, both of them are optimal partitions. 
\end{enumerate}
\qed \end{proof}
%
%

\section{Omitted proof in Section~\ref{sec:tree}}

\let\temp\thelemma
\renewcommand{\thelemma}{\ref{max:ASW:diameter2}}
\begin{lemma}
For tree $T^{\le2}_n$ with diameter at most 2,  $n/2 \le \varphi(T^{\le2}_n) \le 9n/16$ holds. 
\end{lemma}
\let\thelemma\temp
\addtocounter{lemma}{-1}

\begin{proof}
Let $\varphi (K_{1,n-1})$ and $\tilde{\varphi}(K_{1,n-1})$ be the social welfare and the average social welfare of the grand coalition of $K_{1,n-1}$, respectively. Then we can express them as 
$\varphi(K_{1,n-1})={(n-1)(n+2)}/{2n}$ and $\tilde{\varphi}(K_{1,n-1})={(n-1)(n+2)}/{2n^2}$.
Moreover, we differentiate the average social welfare by $n$: ${d\tilde{\varphi}(K_{1,n-1})}/{d{n}}=({4-n})/{2n^3}$.
Therefore, the average social welfare is maximum when $n=4$.
By combining Corollary~\ref{opt:partition}, $n/2 \le \varphi(K_{1,n-1}) \le 9n/16$ holds.
\qed \end{proof}

\let\temp\thelemma
\renewcommand{\thelemma}{\ref{SW:diameter3tree}}
\begin{lemma}
For tree $T^3_n$ with diameter 3, if $k=2$ and $\ell\ge 7$, $k \ge 7$ and $\ell=2$, $k>3$ and $\ell\ge 3$, or $k\ge 3$ and $\ell>3$, $\varphi(T^3_n)<n/2$ holds, and otherwise $\varphi(T^3_n)\ge n/2$.
\end{lemma}
\let\thelemma\temp
\addtocounter{lemma}{-1}

\begin{proof}
For $T^3_n$,
since the utilities of $s_i$, $t_j$, $u_1$ and $u_2$ are $U(s_i, V(T^3_n))=(1+k/2+\ell/3)/n$, $U(t_j, V(T^3_n))=(1+\ell/2+k/3)/n$, $U(u_1, V(T^3_n))=((k+1)+\ell/2)/n$ and $U(u_2, V(T^3_n))=((\ell+1)+k/2)/n$, respectively,  
the social welfare of the grand coalition of $T^3_n$ $\varphi(T^3_n)$ is as follows:
\begin{flalign*}
\varphi(T^3_n)&=\sum_{i=1}^k U(s_i, V(T^3_n))+\sum_{j=1}^{\ell} U(t_j, V(T^3_n))+U(u_1, V(T^3_n))+U(u_2, V(T^3_n))\\
&=\frac{1}{n} \bigl(\ell(1+\frac{\ell}{2}+\frac{k}{3})+k(1+\frac{k}{2}+\frac{\ell}{3})+((k+1)+\frac{\ell}{2})+((\ell+1)+\frac{k}{2})\bigr)\\
&=\frac{1}{n} \bigl (\frac{k^2}{2}+\frac{5k}{2}+\frac{\ell^2}{2}+\frac{5\ell}{2}+\frac{2k\ell}{3}+2 \bigr).
\end{flalign*}
Since $n=k+\ell+2$,
\begin{flalign*}
\varphi(T^3_n)-\frac{n}{2}&=\frac{\frac{k^2}{2}+\frac{5k}{2}+\frac{\ell^2}{2}+\frac{5\ell}{2}+\frac{2k\ell}{3}+2}{k+\ell+2}-\frac{k+\ell+2}{2}\\
&=\frac{k^2+5k+\ell^2+5\ell+\frac{4k\ell}{3}+4}{2(k+\ell+2)}-\frac{(k+\ell+2)^2}{2(k+\ell+2)}\\
&=\frac{1}{2(k+\ell+2)}(k+\ell-\frac{2k\ell}{3})\\
&=\frac{1}{3(k+\ell+2)}\left(\frac{9}{4}-(k-\frac{3}{2})(\ell-\frac{3}{2})\right).
\end{flalign*}
Since $k$ and $\ell$ are positive integers, if $k=2$ and $\ell\ge 7$, $k \ge 7$ and $\ell=2$, $k>3$ and $\ell\ge 3$, or $k\ge 3$ and $\ell>3$, it satisfies that $\varphi(T^3_n)-n/2<0$, and otherwise $\varphi(T^3_n)-n/2\ge0$. 
\qed \end{proof}

\let\temp\thelemma
\renewcommand{\thelemma}{\ref{tree:diameter4}}
\begin{lemma}
For tree $T^4_n$ with diameter 4, $\varphi(T^4_n)\ge{n}/{2}$ holds if \textcolor{black}{$(k,\alpha_k)=(2,2),\break(2,3),(3,2),(4,2)$}, and otherwise $\varphi(T^4_n)<{n}/{2}$ holds.
\end{lemma}
\let\thelemma\temp
\addtocounter{lemma}{-1}

\begin{proof}
For $T^4_n$, since $\alpha_k=\sum_{i=1}^k\ell_i$, the utilities of $v$, $u_i$ and $w_{i, j}$ are denoted by $U(v, V(T^4_n))=k+\alpha_k/2$, $U(u_i, V(T^4_n))=(\ell_i+1)+(k-1)/2+(\alpha_k-\ell_i)/3$ and $U(w_{i, j}, V(T^4_n))=1+\ell_i/2+(k-1)/3+(\alpha_k-\ell_i)/4$, respectively.
Here, let $\beta_k=\sum^{k}_{i=1}\ell_i^2$, then the utilities of $u_i$ and $w_{i, j}$ can be expressed as follows:
{\small
\begin{flalign*}
\sum^k_{i=1}U(u_i, V(T^4_n))&= \sum^k_{i=1}\left((\ell_i+1)+\frac{k-1}{2}+\frac{\alpha_k-\ell_i}{3}\right)=\frac{k\alpha_k}{3}+\frac{2\alpha_k}{3}+\frac{k^2}{2}+\frac{k}{2},\\
\sum^k_{i=1}\sum^{\ell_i}_{j=1}U(w_{i, j},V(T^4_n))&=\sum^k_{i=1}\ell_i \left(1+\frac{\ell_i}{2}+\frac{k-1}{3}+\frac{\alpha_k-\ell_i}{4}\right)=\frac{\alpha_k^2}{4}+\frac{\beta_k}{4}+\frac{k\alpha_k}{3}+\frac{2\alpha_k}{3}.
\end{flalign*}}

Therefore, the social welfare of the grand coalition of $T^4_n$ $\varphi(T^4_n)$ is represented as:
\begin{flalign*}
\varphi(T^4_n)&=\frac{1}{n}\sum_{v\in{V_k}}U(v,V)\\
&=\frac{1}{n}(\sum^k_{i=1}U(u_i, V(T^4_n))+\sum^k_{i=1}\sum^{\ell_i}_{j=1}U(w_{i, j},V(T^4_n))+\textcolor{black}{k+\alpha_{k}/2})\\
&=\frac{1}{n}\left(\frac{\alpha_k^2}{4}+\frac{\beta_k}{4}+\frac{2k\alpha_k}{3}+\frac{11\alpha_k}{6}+\frac{k^2}{2}+\frac{3k}{2}\right).
\end{flalign*}
Since  $n=k+\alpha_k+1$,
{\small
\begin{flalign*}
\varphi(T^4_n)-\frac{n}{2}&=\frac{1}{2(k+\alpha_k+1)}\left(\frac{\alpha_k^2}{2}+\frac{\beta_k}{2}+\frac{4k\alpha_k}{3}+\frac{11\alpha_k}{3}+k^2+3k \right)-\frac{(k+\alpha_k+1)^2}{2(k+\alpha_k+1)}\\
&=\frac{1}{2(k+\alpha_k+1)} \left(\frac{\beta_k}{2}-\frac{\alpha_k^2}{2}-\frac{2k\alpha_k}{3}+\frac{5\alpha_k}{3}+k-1\right).
\end{flalign*}}
Let ${\boldsymbol \ell}=(\ell_1, \ldots, \ell_k)$.
Since $\beta_k-\alpha_k^2=-\sum_{i,j\in[k],i\neq{j}}2\ell_i \ell_j$ and \textcolor{black}{$1/(2(k+\alpha+1))>0$}, if we define $f(k,{\boldsymbol \ell}) =-\sum_{i,j\in[k],i\neq{j}}\ell_i \ell_j-2k\alpha_k/3+5\alpha_k/3+k-1$, then  the sign of  $f(k,{\boldsymbol \ell})$ and the sign of $\varphi(T^4_n)-n/2$ are the same. 
Because $\alpha_k\ge2$ and $\ell_1, \ell_2\ge 1$,
\begin{flalign*}
\frac{\partial f(k,{\boldsymbol \ell})}{\partial{k}}=1-\frac{2\alpha_k}{3}<0.
\end{flalign*}
Moreover, since $k\ge2$ and $\sum_{j\in[k]\setminus \{i\} }\ell_j\ge1$,
\begin{flalign*}
\frac{\partial f(k,{\boldsymbol \ell})}{\partial{\ell_i}}=\frac{5}{3}-\frac{2k}{3}-\sum_{j\in[k]\setminus \{i\} }\ell_j<0.
\end{flalign*}
%
Thus, $f(k,{\boldsymbol \ell})$ is a decreasing function related to $k, \ell_1, \ldots, \ell_k$. Therefore, the case that $f(k, \boldsymbol{\ell})\ge 0$  for $k\ge 2$ and $\ell_1, \ell_2\ge 1$ is \textcolor{black}{$(k,\ell_1, \ell_2, \ell_3, \ldots, \ell_k)=(2,1,1,0, \ldots,  0)$, $(2,1,1,1,\ldots, 0)$, $(3,1,1,0,\ldots, 0)$, $(4,1,1,0,\ldots, 0)$}.
Thus, 
\begin{flalign*}
\varphi(T^4_n)-\frac{n}{2}
\begin{cases}
 \ge0 
&(k,\alpha_k)=(2,2),(2,3),(3,2),(4,2). \\
<0 
&\mbox{otherwise}.
\end{cases}
\end{flalign*} 
\qed \end{proof}
\let\temp\thelemma
\renewcommand{\thelemma}{\ref{tree:diameter6}}
\begin{lemma}
For tree $T^\mu_n$ with diameter $\mu\ge5$, $\varphi(T^\mu_n)<n/2$ holds.
\end{lemma}
\let\thelemma\temp
\addtocounter{lemma}{-1}

\begin{proof}
We show that $\varphi(T^\mu_n)<n/2$ by using mathematical induction. 
We observe that there always exists a path $P_{\mu+1}$ with diameter $\mu$ in $T_n^\mu$. 
By adding vertex and edges to $P_{\mu+1}$ without changing the diameter, we can express any $T^\mu_n$. Then we show that $\varphi(T^\mu_n)<n/2$ in the process of adding vertices and edges.
By Lemma~\ref{lem:pathaverage} and $\varphi(P_6)-(\mu+1)/2=29/10-6/2<0$, the difference between $\varphi(P_{\mu+1})$ and \textcolor{black}{$(\mu+1)/2$} is $\varphi(P_{\mu+1})-(\mu+1)/2<0$ for $P_{\mu+1}$.
We assume that $T^\mu_{k-1}<(k-1)/2$ holds.
Let $u\in{V(T^\mu_{k-1})}$ be a vertex that does not change the diameter of $T^\mu_{k-1}$ even if we connect a new vertex $w$ to $u$ in $T^\mu_{k-1}$.
Moreover, let $T^\mu_k$ be a tree such that a new vertex $w$ is connected to such vertex $u$ and $(w,u)$ are added to $T^\mu_{k-1}$.
\textcolor{black}{
Here, the social welfare of $T^\mu_k$ consists of the social welfare with respect to a new vertex $w$ and the social welfare with in $k-1$ vertices in $T^\mu_{k-1}$. 
The later can be represented as $(k-1)\varphi(T^\mu_{k-1})/k$.
Let $p_k$ be the former one. Then we have $\varphi(T^\mu_k)=(k-1)\varphi(T^\mu_{k-1})/k+p_k$.}

Let $n_2$ and $n_3$ be the number of verticies at distance two and three from $w$ in $T^\mu_k$, respectively, and $n_{\ge4}$ be the number of verticies at distance four and more.
Then $p_k$ is represented as 
\begin{flalign}
p_k=\frac{2+n_2+\frac{2n_3}{3}+\cdots}{k}<\frac{2+n_2+\frac{2n_3}{3}+\frac{n_{\ge4}}{2}}{k}.
\end{flalign}
Moreover, since $k=n_2+n_3+n_{\ge4}+2$ and \textcolor{black}{$\varphi(T^\mu_{k-1})<(k-1)/2$}, it holds that
\begin{flalign}
\frac{k-1}{k}\varphi(T^\mu_{k-1})<\frac{(k-1)^2}{2k}=\frac{(n_2+n_3+n_{\ge4}+1)^2}{2(n_2+n_3+n_{\ge4}+2)}.
\end{flalign}
From \textcolor{black}{inequalities $(1)$ and $(2)$}, it holds that 
\begin{flalign*}
\varphi(T^\mu_k)=\frac{k-1}{k}\varphi(T^\mu_{k-1})+p_k&<\frac{(n_2+n_3+n_{\ge4}+1)^2}{2(n_2+n_3+n_{\ge4}+2)}+\frac{2+n_2+\frac{2n_3}{3}+\frac{n_{\ge4}}{2}}{n_2+n_3+n_{\ge4}+2}\\
&=\frac{(n_2+n_3+n_{\ge4}+1)^2+4+2n_2+\frac{4n_3}{3}+n_{\ge4}}{2(n_2+n_3+n_{\ge4}+2)}.
\end{flalign*}
Finally,
{\small
\begin{flalign*}
\varphi(T^\mu_k)-\frac{k}{2}&<\frac{(n_2+n_3+n_{\ge4}+1)^2+4+2n_2+\frac{4n_3}{3}+n_{\ge4}}{2(n_2+n_3+n_{\ge4}+2)}-\frac{(n_2+n_3+n_{\ge4}+2)^2}{2(n_2+n_3+n_{\ge4}+2)}\nonumber\\ 
&=\frac{4+2n_2+\frac{4n_3}{3}+n_{\ge4}-2n_2-2n_3-2n_{\ge4}-3}{2(n_2+n_3+n_{\ge4}+2)}\nonumber\\
&=\frac{1-\frac{2n_3}{3}-n_{\ge4}}{2(n_2+n_3+n_{\ge4}+2)}\\
&<0.
\end{flalign*}}
Note that since the diameter of $T^\mu_k$ is $\mu\geq5$, it holds that $n_3, n_{\ge4} \geq1$.
\qed \end{proof}
%

\let\temp\thelemma
\renewcommand{\thelemma}{\ref{opt:diameter4tree}}
\begin{lemma}
Any optimal partition of a tree $T$ does not contain a coalition that consists of a tree $T^{\ge4}$ with diameter at least 4.
\end{lemma}
\let\thelemma\temp
\addtocounter{lemma}{-1}

\begin{proof}
We consider coalitions with $(1)$ diameter 4 and $(2)$ diameter at least 5.
\begin{enumerate}
\item

From Lemma~\ref{tree:diameter4}, if \textcolor{black}{$(k,\alpha_k)=(2,2),(2,3),(3,2),(4,2)$}, it holds that $\varphi(T^4)\ge{n/2}$.
For each case, we check whether there is a partition $\mathcal{C}$ which satisfies $\varphi(T^4)<\varphi(T^4,\mathcal{C})$.
\begin{enumerate}
\item If $(k,\alpha_k)=(2,2)$,
we set $\mathcal{C}=\{\{v,u_1,w_{1,1}\},\{u_2,w_{2,1}\}\}$. Since $\varphi(T^4)=77/30<8/3=\varphi(T^4,\mathcal{C})$, the optimal solution of $T$ does not contain $T^4$ as a coalition.

\item \textcolor{black}{If $(k,\alpha_k)=(2,3)$,
we set $\mathcal{C}=\{\{v,u_1,w_{1,1},w_{1,2}\},\{u_2,w_{2,1}\}\}$. Since $\varphi(T^4)=3<10/3=\varphi(T^4,\mathcal{C})$, the optimal solution of $T$ does not contain $T^4$ as a coalition.}

\item If $(k,\alpha_k)=(3,2)$,
we set $\mathcal{C}=\{\{v,u_1,u_2,w_{1,1}\},\{u_3,w_{3,1}\}\}$. Since $\varphi(T^4)=109/36<10/3=\varphi(T^4,\mathcal{C})$, any optimal solution of $T$ does not contain $T^4$ as a coalition.

\item \textcolor{black}{If $(k,\alpha_k)=(4,2)$,
we set $\mathcal{C}=\{\{v,u_3,u_4\},\{u_1,w_{1,1}\},\{u_2,w_{2,1}\}\}$. Since $\varphi(T^4)=7/2<11/3=\varphi(T^4,\mathcal{C})$, the optimal solution of $T$ does not contain $T^4$ as a coalition.}

\end{enumerate}
\item 
Since there exists a partition ${\mathcal C}$ such that  ${\varphi(T^{\ge5},\mathcal{C})}\ge n/2$ by Proposition~\ref{dia2} and $\varphi(T^{\ge5})<n/2$ by Lemma~\ref{tree:diameter6}, 
any optimal solution of $T$ does not contain $T^{\ge5}$ as a coalition.
\end{enumerate}
\qed \end{proof}

\let\temp\thelemma
\renewcommand{\thelemma}{\ref{lem:treemaxSW}}
\begin{lemma}
Let $\mathcal{C}^*$ be the optimal partition of tree $T_n$.
Then, the subgraph $G[C]$ induced by $C\in{\mathcal C}^*$ is one of the following:
\color{black}
$(1)$ a star $K_{1,|C|-1}$ (see Figure~\ref{fig:star}), $(2)$a path $P_4$ of length 3 (see Figure~\ref{fig:P_4}), and $(3)$a tree $T^3_5$ of size five with diameter 3 (see Figure~\ref{fig:P'_4}).
\color{black}
\end{lemma}
\let\thelemma\temp
\addtocounter{lemma}{-1}

\begin{proof}
From Lemma~\ref{opt:diameter4tree}, there is no coalition $C$ in the optimal partition such that the diameter of $G[C]$ is at least $4$.
Thus, we consider diameter at most $2$, and diameter $3$.
\begin{description}
\item[diameter at most 2.]
A tree with diameter at most $2$ is a star.
\item[diameter 3.]
From Corollary~\ref{opt:partition}, we only have to consider a tree of social welfare more than $n/2$.
By Lemma~\ref{SW:diameter3tree}, if $(k,\ell)=(1,\ell'),(2,2),(2,3)$, $(2,4),(2,5),(2,6)$, $(3,3),(3,2)$, $(4,2),(5,2),(6,2)$, $(k',1)$, a tree $T^3$ with diameter 3 satisfies $\varphi(T^3)\ge{|V(T^3)|/2}$.
Here, by using the fact that $k$ and $\ell$ are symmetric, we only consider $(1,\ell')$, $(2,2)$, $(2,3)$, $(2,4)$, $(2,5)$, $(2,6)$ and $(3,3)$.
\begin{description}
\item[Case:$(k,\ell)=(1,\ell')$.]
 \textcolor{black}{From the fact that $\varphi(P_2)=1$ and $\varphi(K_{1,\ell'})=\ell'(\ell'+3)/2(\ell'+1)$}, if $\mathcal{C}=\{\{s_1,u_1\},\{u_2,t_1,\ldots,t_{\ell'}\}\}$,  the social welfare $\varphi(T)$ and $\varphi(T,\mathcal{C})$ are respectively represented as 
\begin{flalign*}
\varphi(T)=\frac{\ell'^2+\frac{19\ell'}{3}+10}{2(\ell'+3)} \mbox{\textcolor{black}{ \ and \ }} \varphi(T,\mathcal{C})=\frac{\ell'^2+5\ell'+2}{2(\ell'+1)}.
\end{flalign*}
Thus, 
\textcolor{black}{
\begin{flalign*}
\varphi(T)-\varphi(T,\mathcal{C})&=\frac{(\ell'+1)(\ell'^2+\frac{19\ell'}{3}+10)-(\ell'+3)(\ell'^2+5\ell'+2)}{2(\ell'+1)(\ell'+3)}\\
&=-\frac{(\ell'-2)}{3(\ell'+1)}.
\end{flalign*}}
\textcolor{black}{If $\ell' \in\{1,2\}$, the social welfare of the grand coalition is actually at least the one of $\mathcal{C}$}. Moreover, if $\ell'\ge3$, the social welfare of $\mathcal{C}$ is more than the grand coalition.
Therefore, candidates for a coalition in the optimal partition are only $T^3_5$ and $P_4$.
\item [Case:$(k,\ell)=(2,2)$.]
If $\mathcal{C}=\{\{s_1,s_2,u_1\},\{u_2,t_1,t_2\}\}$, $\varphi(T^3)=28/9<10/3\break=\varphi(T^3,\mathcal{C})$.
Thus, a tree with diameter three and $(k,\ell)=(2,2)$ is not candidate for coalition in the optimal partition.
\item[Case:$(k,\ell)=(2,3)$.]
If we set $\mathcal{C}=\{\{s_1,s_2,u_1\},\{u_2,t_1,t_2,t_3\}\}$, $\varphi(T^3)=25/7<47/12=\varphi(T^3,\mathcal{C})$ holds.
\item[Case:$(k,\ell)=(2,4)$.]
If we set $\mathcal{C}=\{\{s_1,s_2,u_1\},\{u_2,t_1,t_2,t_3,t_4\}\}$, $\varphi(T^3)=97/24<67/15=\varphi(T^3,\mathcal{C})$ holds.
\item[Case:$(k,\ell)=(2,5)$.]
If we set $\mathcal{C}=\{\{s_1,s_2,u_1\}$, $\{u_2,t_1,t_2,t_3,t_4,t_5\}\}$, \break$\varphi(T^3)=122/27<5=\varphi(T^3,\mathcal{C})$ holds.
\item[Case:$(k,\ell)=(2,6)$.]
If we set $\mathcal{C}=\{\{s_1,s_2,u_1\}$, $\{u_2,t_1,t_2,t_3,t_4,t_5,t_6\}\}$, \break$\varphi(T^3)=5<116/21=\varphi(T^3,\mathcal{C})$ holds.
\item[Case:$(k,\ell)=(3,3)$.]
If we set $\mathcal{C}=\{\{s_1,s_2,s_3,u_1\},\{u_2,t_1,t_2,t_3\}\}$, \break$\varphi(T^3)=4<9/2=\varphi(T^3,\mathcal{C})$ holds.
\end{description}
Since a tree with diameter 3 is $P_4$ if $(k,\ell)=(1,1)$ and $T^3_5$ if $(k,\ell)=(2,1), (1,2)$, candidates for a coalition in the optimal partition are $P_4$ and $T^3_5$
\end{description}
Now, we know that $G[C]$ is one of $(\{v\},\emptyset)$, $K_{1,|C|-1}$, $P_4$ and $T^3_5$ for $C\in\mathcal{C^*}$.
Finally, we show that optimal partition does not contain an isolated vertex $(\{v\},\emptyset)$ as a coalition in a tree.

As with Proposition~\ref{path:isolated}, we assume that an isolated vertex $v$ exists in optimal partition ${\mathcal C}^*$ and lead to contradiction.
In the following, we assume that an isolated vertex $v$ is adjacent to another isolated vertex, the center of a star, one of leaves of a star, a vertex in $P_4$, and a vertex in $T^3_5$.
\begin{description}
\item [An isolated vertex $v$ is adjacent to another isolated vertex $u$.]

It is inconsistent as the social welfare increases by setting partition to ${\mathcal C}^*\setminus \{\{u\}, \{v\}\} \cup \{\{u,v\}\}$.
\item [An isolated vertex $v$ is adjacent to the center of star $K_{1,f}$.]
Since \break$\varphi(K_{1,f+1})=f/2+3/2-1/(f+2)>f/2+1-1/(f+1)=\varphi(K_{1,f})$, the social welfare of ${\mathcal C}^*\setminus (\{V(K_{1,f}),\{v\}\}) \cup (\{V(K_{1,f+1})\})$ is larger than $\mathcal{C}^*$. This is contradiction.
\item [An isolated vertex $v$ is adjacent to one of leaves of star $K_{1,f}$.]
In the case of $f=1, 2$, it is the same as the case of a path. Thus, this is contradiction.
For $f\ge3$, we chose one leaf $u$ in $K_{1,f}$.
Since $\varphi(K_{1,f-1})+\varphi(P_2)=(f^2+3f-2)/2f>(f^2+3f)/2(f+1)=\varphi(K_{1,f})$, if we set $\mathcal{C'}={\mathcal C}^*\setminus (\{V(K_{1,f}),\{v\}\}) \cup (\{V(K_{1,f})\setminus \{u\}, \{u,v\}\})$, $\varphi(G,\mathcal{C'})>\varphi(G,\mathcal{C^*}).$ This is contradiction.
%

\item [An isolated vertex $v$ is adjacent to $P_4$.]
If $v$ is adjacent to a leaf in $P_4$, this is contradiction by Lemma~\ref{path:isolated}.
If $v$ is adjacent not to a leaf, since $\varphi(T^3_5)=8/3>13/6=\varphi(P_4)$, if we set $\mathcal{C'}={\mathcal C}^*\setminus (\{\{V(P_4),\{v\}\}) \cup \{T^3_5\}$, $\varphi(G,\mathcal{C'})>\varphi(G,\mathcal{C^*}).$ This is contradiction.
\item [An isolated vertex $v$ is adjacent to $T^3_5$.]
We consider the following four cases.
\begin{description}
\item[An isolated vertex $v$ is adjacent to  $s_1$.]
Since $\varphi(P_3)+\varphi(P_3)=10/3>8/3=\varphi(T^3_5)$, if we set $\mathcal{C'}={\mathcal C}^*\setminus (\{V(T^3_5),\{v\}\}) \cup (\{V(P_3), V(P_3)\})$, $\varphi(G,\mathcal{C'})>\varphi(G,\mathcal{C^*}).$ This is contradiction.
\item[An isolated vertex $v$ is adjacent to  $u_1$.]
Since $\varphi(P_3)+\varphi(P_3)=10/3>8/3=\varphi(T^3_5)$, if we set $\mathcal{C'}={\mathcal C}^*\setminus (\{V(T^3_5),\{v\}\}) \cup (\{V(P_3), V(P_3)\})$, $\varphi(G,\mathcal{C'})>\varphi(G,\mathcal{C^*}).$ This is contradiction.
\item[An isolated vertex $v$ is adjacent to  $u_2$.]
Since $\varphi(K_{1,3})+\varphi(P_2)=13/4>8/3=\varphi(T^3_5)$, if we set $\mathcal{C'}={\mathcal C}^*\setminus (\{V(T^3_5),\{v\}\}) \cup (\{V(K_{1,3}), V(P_2)\})$, $\varphi(G,\mathcal{C'})>\varphi(G,\mathcal{C^*}).$ This is contradiction.
\item[An isolated vertex $v$ is adjacent to  $t_1$ or $t_2$.]
Since $\varphi(P_2)+\varphi(P_4)\break=19/6>8/3=\varphi(T^3_5)$, if we set $\mathcal{C'}={\mathcal C}^*\setminus (\{V(T^3_5),\{v\}\}) \cup (\{V(P_2), \break V(P_4)\})$, $\varphi(G,\mathcal{C'})>\varphi(G,\mathcal{C^*}).$ This is contradiction.
\end{description}
\end{description}
All the cases contradict the optimality of $\mathcal{C}^*$. Therefore, optimal partition $\mathcal{C}^*$ does not include an isolated vertex.
This completes the proof.
\qed \end{proof}

\section{Omitted proof in Section~\ref{sec:NPh}}~\label{sec:appendix:NPh}
\color{black}
In this section, we prove Theorem~\ref{nphard:4-regular}.
\color{black}
\let\temp\thetheorem
\renewcommand{\thetheorem}{\ref{nphard:4-regular}}
\begin{theorem}
{\sf MaxSWP} is NP-hard even for 4-regular graphs. 
\end{theorem}
\let\thetheorem\temp
\addtocounter{theorem}{-1}

\subsection{Reduction}
In this subsection, we give a reduction from {\sf M3XSAT(3L)}. 
An instance $\psi$ of {\sf M3XSAT(3L)} forms a set of clause, say $S_1,\ldots,S_m$, where each clause consists of three literals from $x_1,x_2,\ldots,x_n$, and each $x_i$ appears three times in $S_j$'s. Note that we can regard a clause as just a $3$-set because of the monotonicity. Since $\sum_{j=1}^m |S_j| = 3m$ and the number of the total occurrences of literals is $3n$, $n=m$ holds. 
From $\psi$, we construct 
a 4-regular graph $G_{\psi}=(V(G_{\psi}), E(G_{\psi}))$ from a given instance of {\sf M3XSAT(3L)}. 
First, we prepare a vertex set $V_{x_i} = \{x^{(1)}_i, x^{(2)}_i, x^{(3)}_i\}$ that corresponds to literal $x_i$. We call these {\em literal vertices}. We also prepare a vertex set $V_{S_j}=\{S^{(1)}_j, S^{(2)}_j\}$ that corresponds to $3$-set $S_j$. These are called {\em clause vertices}. Here, $x^{(1)}_i$, $x^{(2)}_i$ and $x^{(3)}_i$ correspond to the appearances of $x_i$. For example, if $x_4$ appears in $C_2$, $C_3$ and $C_5$, $x^{(1)}_4$, $x^{(2)}_4$ and $x^{(3)}_4$ are associated with $C_2$, $C_3$ and $C_5$, respectively. The vertex set of $G_{\psi}$ consists of $V_{x_i}$'s and $V_{S_j}$'s, that is, $V(G_{\psi}) = \bigcup_{i=1}^n V_{x_i} \cup \bigcup_{j=1}^m V_{S_j}$.  
Thus $|V(G_{\psi})|=3n+2m=5n$ holds. 
We connect these vertices by the following manner: (1) $V_{x_i}$ forms a triangle, that is, there are three edges $(x^{(1)}_i, x^{(2)}_i)$, $(x^{(2)}_i, x^{(3)}_i)$ and $(x^{(3)}_i, x^{(1)}_i)$, (2) vertices in $V_{S_j}$ have an edge, that is, there is an edge $(S^{(1)}_j, S^{(2)}_j)$, and (3) vertices in $V_{x_i}$ are connected with vertices in their corresponding $V_{S_j}$. For example, if $x^{(2)}_i$ appears in $S_j$, there are edges $(x^{(2)}_i, S^{(1)}_j)$ and 
$(x^{(2)}_i, S^{(2)}_j)$ (see Figure \ref{fig:reducedgraph}). We call a subgraph consists of $V_{S_j}$ and its corresponding three literal vertices a {\em clause gadget}. It is easy to see that the degree of every vertex of  $G_{\psi}$ is $4$. 
\begin{figure}[t]
\begin{tabular}{cc}
\begin{minipage}{0.45\hsize}
\centering
\includegraphics[width=55mm, height=48mm]{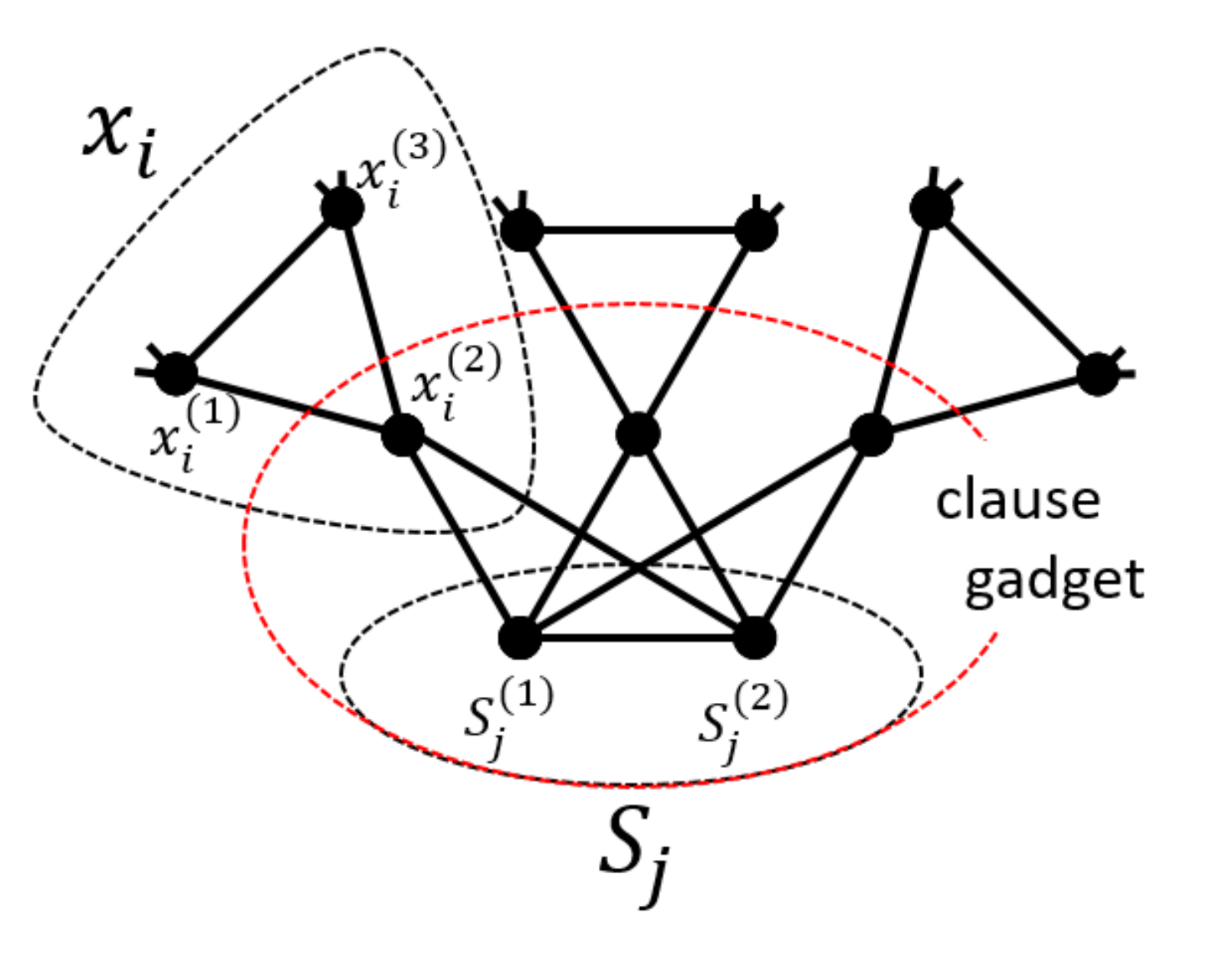}
\caption{Graph $G_{\psi}$}
\label{fig:reducedgraph}
\end{minipage}

\begin{minipage}{0.45\hsize}
\centering
\includegraphics[width=62mm, height=50mm]{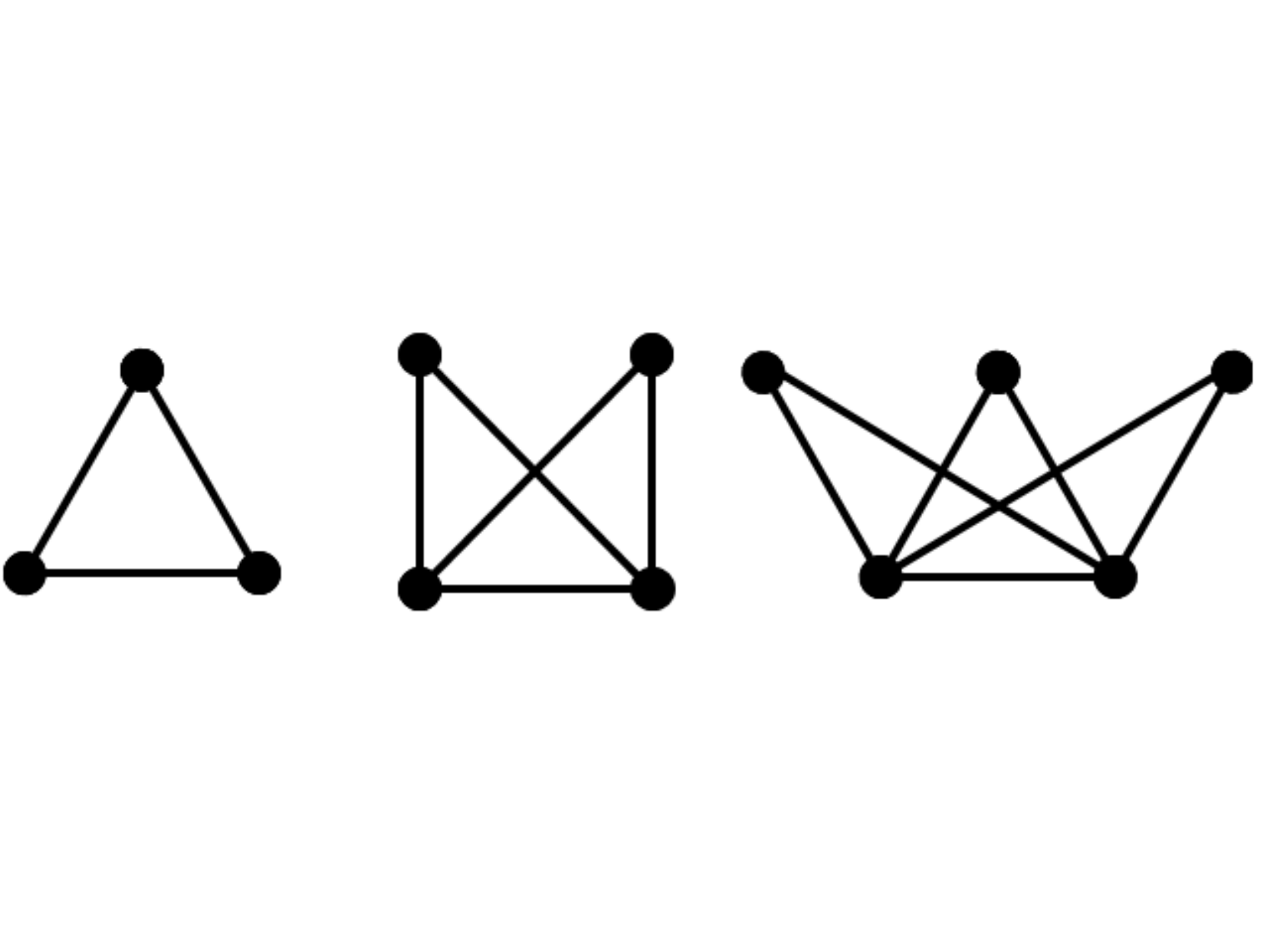}
\caption{(single) triangle, double triangle, and triple triangle}
\label{fig:maxasw}
\end{minipage}
\end{tabular}
\end{figure}

\medskip
\color{black}
We can show the following important property of this $G_{\psi}$. A {\em triangle} is a cycle with $3$ vertices, and a {\em double triangle} is a graph that consists of two triangles sharing an edge, a {\em triple triangle} is a graph that consists of three triangles sharing an edge. See Figure \ref{fig:maxasw}.
\color{black}
\begin{lemma}\label{lem:more2/3}
If a coalition $C$ of $G_{\psi}$ satisfies $\tilde{\varphi}(G[C]) \ge 2/3$, $C$ forms one of the following: triangle, double triangle and triple triangle. The average social welfare of triangle, double triangle and triple triangle are $2/3=0.666\cdots$, $11/16=0.6875$ and $17/25=0.68$, respectively. 
\end{lemma}

To this end, we show several sub-lemmas. The first one is very simple, but it gives an insight of a general property of average social welfare.

\begin{lemma}\label{lem:k}
For a graph $G$ of $k$ vertices and $\ell$ edges, $\tilde{\varphi}(G)\leq(k^2-k+2\ell)/2k^2$ holds. 
\end{lemma}

\begin{proof}\rm
For a vertex $v$, we partition the vertex set $V$ of $G$ as $(V_0,V_1,V_2, \ldots, V_{k})$, where $V_i$ is the set of vertices $i$ distant from $v$. 
Let $d_v$ denote the degree of vertex $v$. Then, $V_0=\{v\}$ and $|V_1|=d_v$. 
By the definition of the utility, we have 
\[
 U(v,G) = \frac{1}{k}\sum_{i=1}\frac{|V_i|}{i} \le \frac{1}{k}(|V_1| + \sum_{i=2}\frac{|V_i|}{2}) = \frac{1}{k}\left(d_v + \frac{k-d_v -1}{2}\right) = \frac{d_v + k -1}{2k},  
\]
and thus the following holds: 
\[
 \tilde{\varphi}(G) \le \frac{1}{2k^2}\sum_{v\in V}(d_v + k -1) = \frac{2\ell + k(k-1)}{2k^2} 
\]
The last equality comes from the handshaking lemma, $\sum_{v\in V}d_v = 2\ell$. 
\qed \end{proof}

By Lemma \ref{lem:k}, we obtain the following. 

\begin{lemma}\label{ASW:degree4}
For a 4-regular graph $G=(V,E)$ with $|V|\geq 9$, the average social welfare of the grand coalition is at most $2/3$, i.e., $\tilde{\varphi}(G)\leq 2/3$ holds. 
Furthermore, for a coalition $C$ that is not a grand coalition of $G$, 
if $|C| \ge 9$, $\tilde{\varphi}(G[C])< 2/3$ holds. 
\end{lemma}
\begin{proof}\rm
By the regularity of $G$, $4|V|=2|E|$ holds. Then by Lemma \ref{lem:k}, 
we have 
\[
 \tilde{\varphi}(G) \le \frac{|V|^2-|V|+4|V|}{2|V|^2} = \frac{1}{2}+\frac{3}{2k} \le \frac{2}{3}. 
\]
For a coalition $C$ that is not a grand, suppose that $G[C]$ is a graph with $k$ vertices and $\ell$ edges. Since 
$G[C]$ contains a vertex whose degree is at most $4$, we have $4k > 2\ell$. 
Thus, by the similar arguement, we have 
\[
\tilde{\varphi}(G[C]) \le \frac{k^2-k+2\ell}{2k^2}<\frac{2}{3}. 
\]
\qed \end{proof}

The next one is not about a general graph but about $G_{\psi}$. 

\begin{lemma}\label{ASW:edge6}
For a coalition $C$ in $G_{\psi}$, $\tilde{\varphi}(G_{\psi}[C])<2/3$ holds if $|C|=6,7,8$. 
\end{lemma}
\begin{proof}\rm
Throughout the proof, we denote $|C|$ by $k$ and the number of edges of $G_{\psi}[C]$ by $\ell$. 

\medskip

($k=6$) 
From Lemma~\ref{lem:k}, \textcolor{black}{if} $k=6$ and $\ell < 9$, $\tilde{\varphi}(G_{\psi}[C])< 2/3$ holds. 
We prove the statement by showing that $G_{\psi}[C]$ contains at most $8$ edges.
We show this by contradiction; we assume that $G_{\psi}[C]$ has $9$ or more edges. Then, the average degree is $\ell \cdot 2/k \ge 3$; it forms a 3-regular graph or a graph with maximum degree is $4$. 
\begin{description}
\item[Case:3-regular graph.] 
Due to $k=6$, $G_{\psi}[C]$ contains at least one clause vertex. If two clause vertices in a clause gadget are contained in $G_{\psi}[C]$, exactly two literal vertices in the same clause gadget should be contained, otherwise it violates $3$-regularity. However, \textcolor{black}{if the degrees of these literal vertices are $3$, the degrees of two extra vertices cannot be $3$}. Thus only one clause vertex in a clause gadget is contained in $G_{\psi}[C]$. Then the three literal vertices in the clause gadget should be contained, but again the degrees of these literal vertices cannot be $3$ by adding two extra vertices. From these, $G_{\psi}[C]$ cannot be a 3-regular graph. 
\item[Case:graph with a vertex of degree 4.] 
We consider the case where $G_{\psi}[C]$ contains a clause vertex of degree $4$ \textcolor{black}{or does not include} a clause vertex of degree $4$.
\begin{description}

\item[Case:including a clause vertex of degree 4.]
It contains all the vertices of a clause gadget, and the remaining one vertex chosen outside the clause gadget.
However, $\ell \leq8$ is obtained for any case.

\item[Case:not including a clause vertex of degree 4.]
The induced subgraph contains three or four vertices in a clause gadget.
Here, if we do not include two clause vertices of the clause gadget, there is no vertex of degree $4$.
For this reason, the induced subgraph contains two clause vertices in the clause gadget.
Here, if the induced subgraph include three literal vertices in the clause gadget, a literal vertex can be of degree $4$.
 Without lose generality, let $v_{x_i}^{(1)}$ be such a literal vertex.
Since the degree of $v_{x_i}^{(1)}$ is 4, other two vertices $v_{x_i}^{(2)}, v_{x_i}^{(3)}\in V_{x_i}$ must be included in the same induced subgraph.

Now, the induced subgraph contains five vertices: two clause vertices and three literal vertices, we only consider the remaining one vertex.
If we choose a clause vertex adjacent to either $v_{x_i}^{(2)}$ or $v_{x_i}^{(3)}$, the number of edges is at most $7$.
If we do not choose a clause vertex adjacent to either $v_{x_i}^{(2)}$ or $v_{x_i}^{(3)}$, the number of edges is at most $8$.
\end{description}
\end{description}


($k=7$)
From Lemma~\ref{lem:k}, if $k=7$ and $\ell\ge12$, $\tilde{\varphi}(G_{\psi}[C])\ge2/3$ holds, but if $k=7$, we show $\ell<12$.
Here we assume that there is a induced subgraph, where $\ell \ge12$. 
The induced subgraph contains a vertex of degree 4 from \textcolor{black}{$12 \cdot 2/7 > 3.4$}.
In the following, we will investigate separately when the induced subgraph contains a clause vertex of degree $4$ or a clause vertex of degree \textcolor{black}{at most} $3$.
\begin{description}
\item [Case:graph with a vertex of degree 4.]
Since the induced subgraph contains all the vertices of a clause gadget, the number of edges is $7$.
Therefore, we consider whether there are vertices where the number of edges is $12$ by including the remaining $2$ vertices.
There are three ways to include vertices, but none of them will exceed the number of edges by more than $12$.
\item [Case:graph with a vertex of degree 3.]
The induced subgraph contains three or four vertices in a clause gadget.
Here, if the induced subgraph does not include two clause vertices of a clause gadget, there is no vertex of degree $4$.
For this reason, the induced subgraph includes two clause vertices within a clause gadget.
If a coalition contains three vertices in a clause gadget, the vertex of degree $4$ is the literal vertex in a clause gadget.
That is, since the coalition contains all vertices adjacent to the literal vertex, we consider only the remaining two vertices.
It may be possible to include two vertices in other gadget adjacent to literal vertices of degree $4$, but the number of edges is at most $9$. 
If the induced subgraph includes four vertices in a clause gadget, two clause vertices are included as well as including the three vertices in a clause gadget.
Therefore, we only consider the remaining one vertex.
In this case, there are two combinations, but in either case the number of edges is at most $9$.
\end{description}


($k=8$)
From Lemma~\ref{lem:k}, if $k=8$ and $\ell\ge15$, $\tilde{\varphi}(G_{\psi}[C])\ge2/3$ holds, but if $k=8$, we show $\ell<15$.
we assume that the induced subgraph satisfy $\ell\ge15$.
Since the $G_{\psi}$ is two-connected four-regular, the degree of the subgraph is at most $4$ and includes at least two vertices of the degree of \textcolor{black}{at most} $3$. 
Therefore, the maximum edges of the subgraph with the number of eight vertices is $15$ from $((8-2) \cdot 4+2\cdot3)/2=15$, but we show that there is no subgraph that realizes this number of edges.
Suppose that $G[C]$ includes six vertices of degree $4$ and two vertices of degree $3$.
Here, we consider the case where the induced subgraph contains a clause vertex of degree $4$, or a clause vertex of degree \textcolor{black}{at most} $3$.
\begin{description}
\item [Case:graph with a vertex of degree 4.]
The induced subgraph contains all the vertices in a clause gadget.
Therefore, we think about the remaining three vertices.
In order to make the degree of literal vertex in a clause gadget be equal to or lager than $3$, the induced subgraph includes literal vertex of other gadget, but we can not make degree of literal vertex of other gadget $3$ or more. 
This is contradiction.
\item [Case:graph with a vertex of degree 3.]
The induced subgraph includes two clause vertices and two literal vertices in a clause gadget.
Here, the degree of the clause vertex is $3$, that is, the vertices other than the clause vertices are degree $4$.
However, in order to make degree of literal vertex in a clause gadget $4$, the induced subgraph need include literal vertex of other gadget.
Moreover, in order to make degree of literal vertex in a clause gadget $4$, the induced subgraph need include clause vertex.
This is contradiction.
\end{description}\qed \end{proof}

\begin{lemma}
For any coalition $C\in\mathcal{C}$ of $G_{\psi}$, $\tilde{\varphi}(G_{\psi}[C])\leq11/16$ holds.
\end{lemma}
\begin{proof}
From Lemmas \ref{ASW:degree4} and \ref{ASW:edge6}, if $k\ge6$, $\tilde{\varphi}(G_{\psi}[C])\leq2/3$ holds.
We consider the average social \textcolor{black}{welfare} for $k=2,3,4,5$.
If $k=2$, the upper bound of the average utility is $1/2$.
Moreover, if $k=3$, the average social welfare of a complete graph of size $3$ is maximum and $2/3$.
If $k=4$, the average social welfare of a clique of size 4 minus one edge is maximum in $G_{\psi}$ since it is  two-connected four-regular and it  does not contain  a clique of size 4.
In this case, the average social welfare of  a clique of size 4 minus one edge is $11/16$.
\textcolor{black}{Since the average social welfare on other subgraphs in $G_{\psi}$ of size $4$ is at most $11/16$, maximum average social welfare of subgraph in $G_{\psi}$ of size $4$  is also $11/16$.}
%
If $k=5$, there are 7 induced subgraphs of $G_{\psi}$. 
Since the upper bound of the average utility is proportional to the number of edges from Lemma~\ref{lem:more2/3}, we only consider the case of the largest number of edges.
Such an induced subgraph is a triple triangle corresponding to a clause gadget. 
The average social welfare of a triple triangle is $17/25$.
\textcolor{black}{Since the average social welfare on other subgraphs in $G_{\psi}$ of size $5$ is at most $17/25$, maximum average social welfare of subgraph in in $G_{\psi}$ of size $5$ is also $17/25$.}
Therefore, for any coalition $C\in\mathcal{C}$ of $G_{\psi}$, $\tilde{\varphi}(G_{\psi}[C])\leq11/16$ holds.

\qed \end{proof}

\subsection{NP-hardness of {\sf MaxSWP}}
From Lemma~\ref{lem:more2/3}, we prove the following lemma. 
\begin{lemma}\label{NPhard}
An instance $\psi$ of {\sf M3XSAT(3L)} is a yes-instance if and only if 
$G_{\psi}$ has a partition $\mathcal{C}$ such that $\varphi(G_{\psi},\mathcal{C})=41n/12$.
\end{lemma}

\begin{proof}
($\Rightarrow$) Assume that there is a truth assignment, and we construct a partition of $G_{\psi}$ from the assignment. In the partition, each $V_{x_i}$ with true $x_i$ forms a coalition \textcolor{black}{as a triangle}, and each $V_{S_j}$ \textcolor{black}{together with vertices of false literals} forms a coalition \textcolor{black}{as a double triangle}.  For example, suppose that $S_j=\{x^{(1)}_1,x^{(1)}_2,x^{(1)}_3\}$ and $S_j$ is satisfied by $x^{(1)}_1$ (actually $x_i$). We then consider a coalition $\{S^{(1)}_j,S^{(2)}_j,x^{(1)}_2, x^{(1)}_3\}$ for $V_{S_j}$ which forms a double triangle.
Since the truth assignment satisfies all the clauses, each $V_{S_j}$ is a part of coalition of size $4$. Literal vertices $V_{x_i}$ themselves form a coalition if $x_i=1$, and otherwise they are included in $C_j$'s. 
Since the utility of the coalition of a double triangle is $11/4$ and the utility of a triangle is $2$, the utility of the reduced graph is $\varphi(G_{\psi},\mathcal{C})$=$11n/4+2\cdot(n/3)=41n/12$.
%

\noindent 
($\Leftarrow$) Assume that there exists a partition $\mathcal{C}$ of $G_{\psi}$ whose social welfare is at least $41n/12$, and the average social welfare is at least $(41n/12)/(5n)=41/60=0.683\cdots$. This and Lemma \ref{lem:more2/3} imply that $\mathcal{C}$ contains at least one double triangle.  
Notice that a coalition can form a double triangle only in a clause gadget. If a clause gadget contains a coalition of a double triangle, one literal vertex is left. \textcolor{black}{Since such a literal vertex belongs to a coalition of a subgraph of a triangle, its utility is at most $2/3$.} Again, by Lemma \ref{lem:more2/3}, the average social welfare of a vertex not included in double triangle is at most $17/25$. From these, if $\mathcal{C}$ contains $p$ double triangles, the social welfare is at most 
\[
 \frac{11}{16}\cdot 4p + \frac{2}{3} \cdot p + \frac{17}{25}\cdot 5(n-p) = \frac{17}{5} n + \left(\frac{41}{12} - \frac{17}{5} \right)p.  
\]
Since the social welfare of $\mathcal{C}$ is at least $41n/12$ by the assumption, we have
\[
 \frac{17}{5} n + \left(\frac{41}{12} - \frac{17}{5} \right)p \ge \frac{41n}{12}, 
\]
and thus $p\ge n$ holds. On the other hand, $p$ is the number of double triangles, and it is at most $n$. It follows that $p=n$ and every clause gadget contains a double triangle as a coalition in $\mathcal{C}$. Furthermore, a literal vertex not included in a double triangle belongs to a coalition of a triangle, which corresponds to a literal $x_i$. Then by assigning true to such $x_i$, every clause gadget includes exactly one true literal, which is a solution of {\sf M3XSAT(3L)}. 
\qed \end{proof}

\color{black}
By this Lemma~\ref{NPhard} and the NP-hardness of {\sf M3XSAT(3L)}, we complete the proof of Theorem~\ref{nphard:4-regular}.
\color{black}

%

\end{document}